\documentclass{jfm}
\usepackage{graphicx}
\usepackage{epstopdf, epsfig}
\usepackage{amsfonts,amssymb,amsmath,mathbbol}   

\usepackage{xcolor} 

\newcommand{\ks}{\textcolor{black}} 
\newcommand{\som}{\textcolor{magenta}} 
\def\We{{\it We}}

\shorttitle{Droplet size distribution under swirl flow}
\shortauthor{S. S. Ade, P. K. Kirar, L. D. Chandrala and K. C. Sahu}

\title{Droplet size distribution in a swirl airstream using in-line holography technique}

\author{Someshwar Sanjay Ade,\aff{1}
Pavan Kumar Kirar,\aff{2}
Lakshmana Dora Chandrala\aff{3}
 \and Kirti Chandra Sahu\aff{2}\corresp{\email{lchandrala@mae.iith.ac.in, ksahu@che.iith.ac.in}}}

\affiliation{\aff{1} Center for Interdisciplinary Program, Indian Institute of Technology Hyderabad, Kandi - 502 284, Sangareddy, Telangana, India
\aff{2}Department of Chemical Engineering, Indian Institute of Technology Hyderabad, Kandi - 502 284, Sangareddy, Telangana, India
\aff{3}Department of Mechanical and Aerospace Engineering, Indian Institute of Technology Hyderabad, Kandi - 502 284, Sangareddy, Telangana, India}

\begin{document}

\maketitle

\begin{abstract}
We investigate the morphology and size distribution of satellite droplets resulting from the interaction of a freely falling water droplet with a swirling airstream of different strengths by employing shadowgraphy and deep-learning-based digital in-line holography techniques. We found that the droplet exhibits vibrational, retracting bag and normal breakup phenomena for the no swirl, low and high swirl strengths for the same aerodynamic field. In the high swirl scenario, the disintegrations of the nodes, rim, and bag-film contribute to the number mean diameter, resulting in smaller satellite droplets. In contrast, in the low swirl case, the breakup of the rim and nodes only contributes to the size distribution, resulting in larger droplets. The temporal variation of the Sauter mean diameter reveals that for a given aerodynamic force, a high swirl strength produces more surface area and surface energy than a low swirl strength. The theoretical prediction of the number-mean probability density of tiny satellite droplets under swirl conditions agrees with experimental data. However, for the low swirl, the predictions differ from the experimental results, particularly due to the presence of large satellite droplets. Our results reveal that the volume-weighted droplet size distribution exhibits two (bi-modal) and three (multi-model) peaks for low and high swirl strengths, respectively. The analytical model that takes into account various mechanisms, such as the nodes, rim, and bag breakups, accurately predicts the shape and characteristic sizes of each mode for the case of high swirl strength. 
\end{abstract}



\section{Introduction} \label{sec:intro}
Raindrops reach earth in a variety of shapes and sizes due to their complex interactions with the atmosphere and accompanying microphysical processes, such as fragmentation \citep{Villermaux2009single,kostinski2009raindrops}, coalescence \citep{schlottke2008direct,chaitanya2021study} and phase change \citep{schlottke2008direct}. These microphysical processes are further influenced by various factors, including meteorological conditions, cloud type from which raindrops originate, the topology of the earth, and air movement in the atmosphere. The distribution of shape and size of raindrops is one of the important factors in rainfall modelling. This was first noticed by \cite{bentley1904studies,von1904regen}. Subsequently, \cite{marshall1948distribution} established a relationship that correlates the average diameter of raindrops with rainfall rate, which has been used in rainfall modelling even today \citep{Villermaux2009single}. Although several researchers have studied the droplet fragmentation phenomenon due to its importance in a variety of industrial applications, such as combustion, surface coating, pharmaceutical manufacturing, and disease transmission modelling \citep{villermaux2007fragmentation,marmottant2004spray,soni2020deformation,jackiw2021aerodynamic,kirar2022experimental,xu2022droplet}, there have been very few studies on the size distribution of satellite droplets after fragmentation of a large droplet \citep{srivastava1971size,Villermaux2009single,jackiw2022prediction,gao2022observation}. Moreover, to our knowledge, no one has previously investigated the droplet size distribution under swirl airstream, even though such a situation is frequently encountered during rainfall \citep{lewellen2000influence,haan2017critical} and in many industrial applications \citep{soni2021liquid,candel2014dynamics}. This is the subject of the present investigation.

In a continuous airstream, a droplet undergoes morphological changes due to the development of the Rayleigh–Taylor instability during its early inflation stage and later encounters fragmentation due to the Rayleigh–Plateau capillary instability \citep{taylor1963shape},\som{} and nonlinear instability of liquid ligaments \citep{jackiw2021aerodynamic,jackiw2022prediction}. The droplet exhibits different breakup modes, such as vibrational, bag, bag-stamen, multi-bag, shear, and catastrophic breakup modes, depending on the speed and direction of the airstream \citep{pilch1987use,dai2001temporal,cao2007new,guildenbecher2009secondary,suryaprakash2019secondary,soni2020deformation}. The Weber number, defined as $\We \equiv \rho_a U^2 d_0/\sigma$, wherein $\rho_a$, $\sigma$, $U$ and $d_0$ denote the density of the air, interfacial tension, average velocity of the airstream and equivalent spherical diameter of the droplet, is used to characterised the droplet breakup phenomenon. A droplet exhibits shape oscillations at a specific frequency for low Weber numbers and fragments into small droplets of comparable size. This is known as the vibrational breakup. As the Weber number is increased, the droplet evolves to form a single bag on the leeward side, encircled by a thick liquid rim. As a result of the bag and rim fragmentation, tiny and slightly bigger droplets are formed. This process is known as the bag breakup phenomenon. The bag-stamen and multi-bag morphologies are identical to the bag breakup mode, but with a stamen formation in the drop's centre, resulting in a large additional drop during the breakup (bag-stamen fragmentation) and several bags formation (multi-bag mode fragmentation). In shear mode, the drop's edge deflects downstream, fracturing the sheet into small droplets. When the Weber number is exceedingly high, the droplet explodes into a cluster of tiny fragments very quickly, resulting in catastrophic fragmentation. \cite{taylor1963shape} was the first to find the critical Weber number $(\We_{cr})$ at which the transition from the vibrational to the bag breakup occurs. Subsequently, several researchers have investigated droplet fragmentation subjected to an airstream in cross-flow configuration (when the droplet interacts with the airstream in a direction orthogonal to gravity) and in-line configurations, such as co-flow and oppose-flow, when the droplet interacts with the airstream flowing along and opposite to the direction of gravity, respectively. The critical Weber numbers in cross-flow \citep{pilch1987use,guildenbecher2009secondary,krzeczkowski1980measurement,Jain2015,hsiang1993drop,wierzba1990deformation,kulkarni2014bag,wang2014} and oppose-flow \citep{Villermaux2009single,villermaux2011distribution} configurations were found to be about 12 and 6, respectively. \cite{soni2020deformation} showed that in the case of an oblique airstream, a freely falling droplet exhibits curvilinear motion while undergoing topological changes and that the critical Weber number decreases when the direction of the airstream changes from the cross-flow to the oppose-flow condition. The value of $\We_{cr}$ was found to be about 12 in the cross-flow configuration, and its value approaches 6 when the inclination angle of the airstream with horizontal is greater than $60^\circ$. The initial droplet size, fluid properties, ejection height from the nozzle, and velocity profile/potential core region were also known to affect the critical Weber number \citep{hanson1963,wierzba1990deformation}. All these investigations are for straight airflow in the no-swirl condition. The deformation and breakup of a water droplet falling in air have also been investigated \citep{szakall2009wind,agrawal2017nonspherical,balla2019shape,agrawal2020experimental}. 

\ks{A few researchers \citep{merkle2003effect,rajamanickam2017dynamics,kumar2019large,patil2021air,soni2021liquid} have examined the characteristics of swirl flow using high-speed imaging and particle image velocimetry (PIV) techniques.} The droplet morphology and breakup phenomenon in a swirling airflow is more complex than in a straight airflow without a swirl. In the case of swirl airflow, the droplet encounters oppose-flow, cross-flow, and co-flow situations simultaneously, depending on its ejection position, airstream velocity, and swirl strength. Recently, \cite{kirar2022experimental} experimentally analysed the droplet fragmentation phenomenon under a swirling airstream using shadowgraphy and particle image velocimetry techniques. They discovered a new breakup mode, termed as `retracting bag breakup', due to the differential flow field created by the wake of the swirler's vanes and the central recirculation zone in swirl airflow. A theoretical analysis based on the Rayleigh-Taylor instability was also developed that uses a modified stretching factor than the straight airstream to predict the droplet deformation process under a swirl airstream. \cite{apte2009stochastic} performed large-eddy simulations of spray atomization from a swirl injector. They used a Pratt and Whitney injector (a high-bypass turbofan engine family) to create a spray alteration/rotation. They found that the number probability density of the atomized droplets follows the Fokker-Planck equation. This partial differential equation provides the temporal evolution of the probability distribution of the velocity of a particle subjected to drag and random forces similar to that in Brownian motion. Apart from the studies mentioned above, \cite{shanmugadas2018characterization} experimentally investigated wall filming and atomization from a swirl cup injector. The injector uses a simplex nozzle and a primary swirler to create a liquid rim and a secondary swirler to create the ligaments and droplets from the rim by shearing action. Their study reveals that the atomization/fragmentation process is a strong function of the central recirculation zone of the primary swirl. Further, \cite{shao2018sheet} performed direct numerical simulations on the swirling sheet breakup process leading to the formation of sheets, ligaments, and droplets. They found that the size distribution of droplets in a swirling sheet breakup process approximately follows the log-normal distribution. As this discussion indicates, despite significant research on the various droplet breakup processes and flow characteristics in straight and swirling airstreams, the size distribution of satellite droplets following the fragmentation of a single droplet has not been investigated yet. 

The size distribution of satellite droplets due to fragmentation is commonly analysed using Laser Diffraction (LD), Phase Doppler Particle Analyzer (PDPA), and in-line holography techniques \citep{gao2013quantitative,katz2010applications,kumar2019automated}. In the LD technique, a radial sensor measures the light scattered by the satellite droplets in the forward direction. This is not an imaging-based approach and must be used in conjunction with a model-based inversion to estimate the size distribution function of the detected droplets. The PDPA is a single-point measurement technique where the phase difference of scattered light captured at multiple angles is utilized to obtain the size of the droplets. The LD and PDPA techniques are suitable for applications, such as spray atomization, where droplets are produced continuously. However, both methods suffer from a number of limitations, such as small sampling volume in the PDPA method and lack of spatial resolution in the LD technique. Moreover, while the PDPA is restricted to spherical particles, the LD technique is not sensitive to single droplet analysis. 

In contrast, the digital in-line holography technique, which employs a deep-learning-based image processing method, has recently emerged as a powerful tool for capturing three-dimensional information about an object with a high spatial resolution (\cite{shao2020machine}). This technique can also offer the spatial distribution of satellite droplets. Thus, it is a superior choice to be used in the current study to analyse the size distribution of satellite droplets resulting from the fragmentation of a water droplet subjected to a swirl airstream. A few researchers have employed in-line holography to determine the droplet's size distribution in a straight airstream in no-swirl conditions. For instance, for an ethanol droplet in a cross-flowing airstream, \cite{gao2013quantitative,guildenbecher2017characterization} used the in-line holography technique to obtain the three-dimensional position and velocities of fragments in the bag breakup and shear-thinning regimes. \cite{li2022secondary} used this technique to analyse the size distribution of satellite droplets under an induced shock wave. The investigation by \cite{jackiw2022prediction} is one of the most relevant studies in the context of the present work. \cite{jackiw2022prediction} developed an analytical model for the combined multi-modal distribution resulting from the aerodynamic breakup of a droplet under no-swirl conditions. They demonstrated that their theoretical predictions agree well with the experimental results of \cite{guildenbecher2017characterization}. Although \cite{jackiw2022prediction} performed fragmentation experiments for a water droplet under straight airflow without swirling using a shadowgraphy technique, they only analysed the size distributions of satellite ethanol droplets reported by \cite{guildenbecher2017characterization} using a digital in-line holography technique.  

In the present work, we investigate the size distribution of satellite droplets resulting from the fragmentation of a freely falling water droplet under a swirl airstream of different strengths by employing both shadowgraphy and digital in-line holography techniques. We found that while the satellite droplets show mono-modal size distribution for the no-swirl airstream, the fragmentation exhibits bi-modal and multi-modal distributions in a swirl flow for the low and high swirl strengths, respectively. A theoretical analysis accounting for various mechanisms, such as the nodes, rim, and bag breakup mode, is also carried out to determine the size distribution of droplets for different swirl strengths. To the best of our knowledge, the present study is the first attempt to estimate the size distribution of satellite droplets under swirl airstreams, which are commonly observed in many natural and industrial applications.

The rest of the paper is organized as follows. In \S\ref{sec:expt}, the experimental set-up and procedure for both the shadowgraph and digital in-line holography are elaborated. In order to demonstrate the digital in-line holography technique used in our work, we have considered a typical bag fragmentation of a water droplet under a straight airstream without swirl and compared our result with the size distribution of an ethanol droplet considered by \cite{guildenbecher2017characterization}. The influence of different swirl airstreams on the droplet fragmentation and the resulting size distribution are discussed in \S\ref{sec:dis}. A theoretical model is employed to predict the size distributions of satellite droplets obtained from our in-line holography technique. The concluding remarks are given in \S\ref{sec:conc}.

\section{Experimental set-up}
\label{sec:expt}

\begin{figure}
\centering
\includegraphics[width=0.9\textwidth]{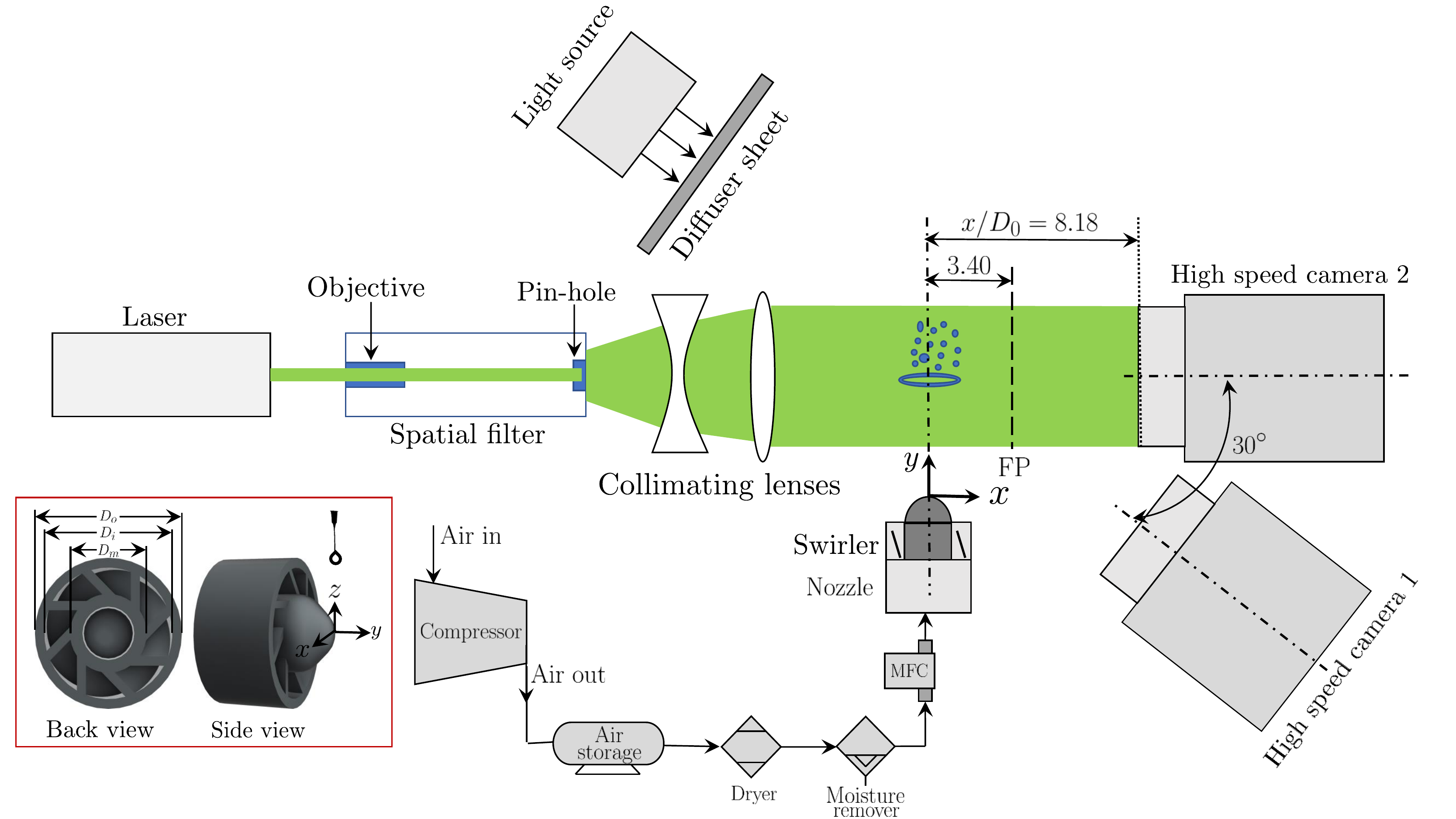}
\caption{Schematic of the experimental set-up (top view) equipped with a shadowgraphy and digital in-line holography systems to investigate the size distribution of satellite droplets due to the fragmentation of a freely falling water droplet under an imposed swirl airstream. The inset depicts the back and side views of a swirler. Here, FP and MFC represent the focal plane associated with the high-speed camera 2 and mass flow controller, respectively. A picture of our experimental set-up is presented as Figure S1 in the supplementary information.}
\label{fig1}
\end{figure}

In the present study, while a shadowgraphy technique is employed to investigate the droplet deformation and breakup phenomena of a freely falling droplet under an imposed swirl airstream, a digital in-line holography technique is used to analyze the size distribution of satellite droplets after fragmentation of the primary drop. The experimental set-up consists of (i) an air nozzle (18 mm diameter) with a swirler, (ii) a droplet dispensing needle connected to a three-dimensional (3D) traverse system, (iii) a continuous wave laser, (iv) a spatial filter arrangement, (v) collimating optics which include concave and convex lenses, (vi) two high-speed cameras and (vii) a diffused backlit illumination. A schematic diagram of the complete experimental set-up is shown in figure \ref{fig1}. The inset at the bottom of figure \ref{fig1} shows the back and side views of the swirler. To change the strength of the swirl airstream, two different types of eight-vane swirlers are used, one with a vane angle, $\theta=30^\circ$ and the other with $\theta=45^\circ$. The rest of the dimensions of the swirlers are the same: outer diameter ($D_o=22$ mm), inner diameter ($D_i=20$ mm), dome diameter ($D_m=12$ mm) and blade thickness ($=1$ mm). The swirl strength is characterized by the geometric Swirl number $(Sw)$, which is defined as \citep{beer1974combustion}
\begin{equation}
Sw= \frac{2}{3} \left( \frac{ 1- \left (D_i/D_o \right)^3 } {1- \left (D_i/D_o \right)^2} \right)\tan\theta,
\end{equation}
such that when $\theta=30^\circ$ and $45^\circ$, we achieve $Sw=0.47$ (low swirl strength) and $Sw=0.82$ (high swirl strength), respectively. We have also calculated the flow-based swirl number $(Sw_f)$ using the stereo-PIV measurements \citep{kirar2022experimental}. The flow swirl number $(Sw_f)$ is given by 
\begin{equation}
Sw_f = \frac{\int_{0}^{R} u_y u_{\theta} r^2 dr}{R \int_{0}^{R} {u_y}^2  r dr},
\end{equation}
where $u_y$ and $u_{\theta}$ are the axial and tangential components of the velocity, and $R$ denotes the radius of the swirler \citep{candel2014dynamics}. We found that the values of $Sw_f$ are about 0.42 and 0.58 for the corresponding geometric swirl numbers $Sw=0.47$ and 0.82, respectively.

The swirlers are fabricated with a tough resin using 3D printing technology and have provision for attaching at the exit of a metallic circular nozzle. In order to remove inlet airstream disturbances and straighten the flow, a honeycomb pallet is placed upstream of the nozzle. The air nozzle is connected to an ALICAT digital mass flow controller (model: MCR-500SLPM-D/CM, Make: Alicat Scientific, Inc., USA), which can regulate flow rates between 0 and 500 standard litres per minute. The accuracy of the flow metre is about 0.8\% of the reading + 0.2\% of the full scale. The mass flow controller is coupled to an air compressor for air supply. An air dryer and moisture remover are installed in the compressed airline to maintain dry air during the experiments. 

A Cartesian coordinate system $(x,y,z)$ with its origin at the center of the swirler tip, as shown in figure \ref{fig1}, is used to analyze the dynamics. A syringe pump (model: HO-SPLF-2D, Make: Holmarc Opto-Mechatronics Pvt. Ltd., India) connected to a dispensing needle is used to create consistent-sized droplets of water (spherical diameter, $d_0 = 3.09 \pm 0.07$ mm). We use a 20 Gauge needle to dispense the droplets. The outer and inner diameters of the needle are 0.908 mm and 0.603 mm, respectively, and the length of the needle is 25.4 mm. The dispensing needle is fixed to a three-dimensional traverse mechanism that allows it to change the location of its tip $(x_d,y_d,z_d)$ and thus control the droplet's location in the swirl airstream. A flow rate of 20 $\mu$l/s is maintained in the syringe pump to create a water droplet at the tip of a blunt needle. This flow rate is low enough that only gravity can detach the droplets from the needle. We ensure that no other droplets emerge from the needle during the droplet's interaction with the swirl flow airstream. 

For shadowgraphy, a high-speed camera (model: Phantom VEO 640L, make: Vision Research, USA) with a Nikkor lens of a focal length of 135 mm and a minimum aperture of $f/2$  is employed. It is designated as high-speed camera 1 in the schematic diagram (figure \ref{fig1}). The high-speed camera 1 is positioned at $x=180$ mm while maintaining an angle of $-30^\circ$ with the $x$ axis. To illuminate the background uniformly, a high-power light-emitting diode (model: MultiLED QT, Make: GSVITEC, Germany) is used along with a diffuser sheet as shown in figure \ref{fig1}. The images captured using the high-speed camera 1 have a resolution of $2048 \times 1600$ pixels, and they are recorded at 1800 frames per second (fps) with an exposure duration of 1 $\mu$s and a spatial resolution of 29.88 $\mu$m/pixel. The droplet's image sequence is stored in the internal memory of the camera and then transferred to a computer for further processing.  

The digital in-line holography set-up consists of a laser, spatial filter, collimating lenses, and the high-speed camera 2 (model: Phantom VEO 640L; make: Vision Research, USA) with a Tokina lens (focal length of $100$ mm and a maximum aperture $f/2.8$, model: AT-X M100 PRO D Macro). The high-speed camera 2 is positioned at $x=180$ mm as shown in figure \ref{fig1}. A laser beam is generated using a continuous wave laser (model: SDL-532-100T, make: Shanghai Dream Lasers Technology Co. Ltd.) that produces 100 mW output power at a wavelength of 532 nm. The laser beam is passed through a spatial filter to produce a clean beam. The spatial filter consists of an infinity-corrected plan achromatic objective (20X magnification, make: Holmarc Opto-Mechatronics Ltd.) and a 15 $\mu$m pin-hole. The beam from the spatial filter is expanded using a plano-concave lens and then collimated using a plano-convex lens (make: Holmarc Opto-Mechatronics Ltd.) before it illuminates the droplet field of view. The diameter of the collimated beam is 50 mm. The interference patterns created due to the droplet's disintegration are recorded using the high-speed camera 2 with a resolution of $2048 \times 1600$ pixels at 1800 fps, at an exposure duration of 1 $\mu$s, and a spatial resolution of 19.27 $\mu$m/pixel. The high-speed camera 1 (used for the shadowgraphy) and high-speed camera 2 (employed for the digital in-line holography) are synchronized using a digital delay generator (model: 610036, Make: TSI, USA). While presenting the results of droplet size distributions, the error bars display the standard deviation of three repetitions for each set of parameters. 

\begin{figure}
\centering
\includegraphics[width=0.9\textwidth]{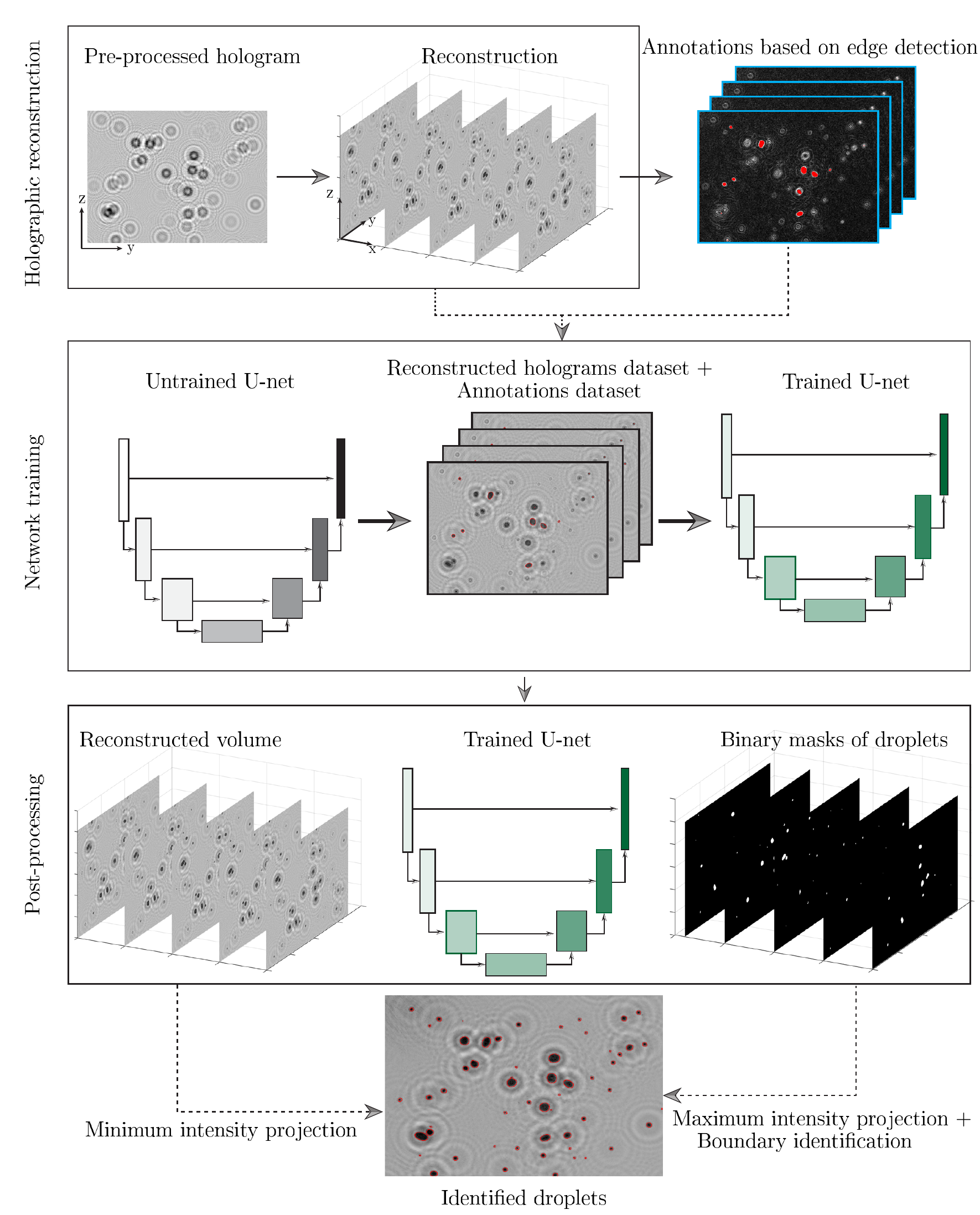}
\caption{Various steps of the digital in-line holography technique employed in the present study.}
\label{fig2b}
\end{figure}

\subsection{Digital in-line holography technique}

Digital in-line holography uses a collimated, coherent, and expanded laser beam. The first step in digital holography is to record the interference patterns created by the scattered light from the droplets (object wave) and the unscattered background illumination (reference wave) on a camera sensor. Therefore, the recorded hologram contains both the amplitude and phase information of the object wave. The second step deals with the reconstruction of the hologram. The hologram is numerically illuminated in the reconstruction process with a reference beam to obtain droplet information at different depths. In in-line holography, the single beam acts as both a reference  and an object beam.

\begin{figure}
\centering
\includegraphics[width=0.95\textwidth]{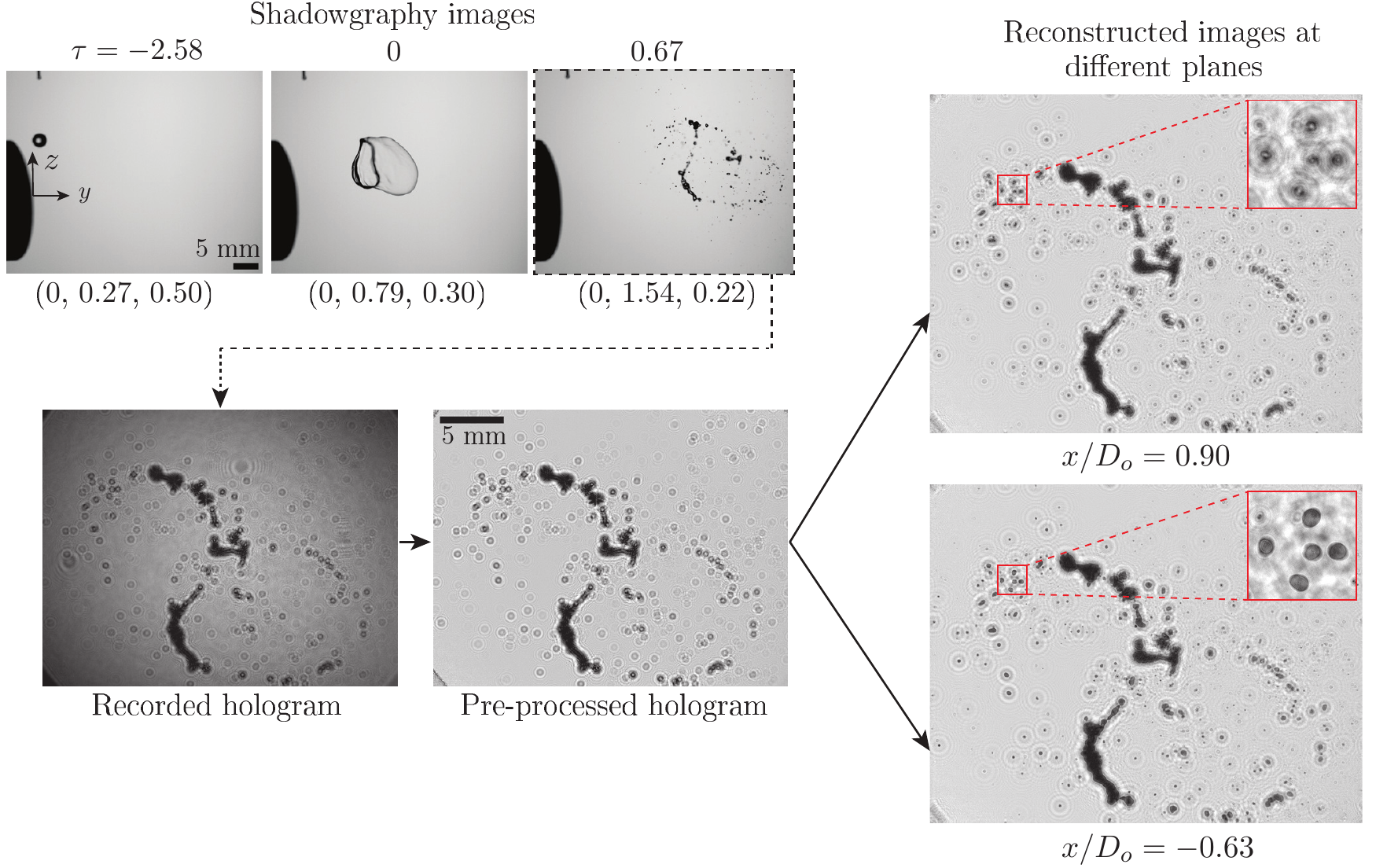}
\caption{A demonstration of different steps to obtain the droplet size distribution after breakup using the in-line holography technique. The temporal evolution of the bag formation and breakup obtained using the shadowgraphy technique is shown in the top left panel. The dimensionless coordinate of the droplet ($x/D_{0}$, $y/D_{0}$, $z/D_{0}$) is mentioned at the bottom of the shadowgraphy images at different instants. The outlet of the nozzle and the tip of the dispensing needle are indicated in the left and top of the shadowgraphy images. The image at $\tau=0.67$ is considered to reconstruct the hologram at different planes in the $x$ direction. Here, $\We=14.53$, $Sw = 0$ and $(x_d/D_o,y_d/D_o,z_d/D_o) = (0.0, 0.27, 0.95)$. An animation showing the different holography reconstructed planes at $\tau=0.67$ is provided as supplementary movie 1.}
\label{fig3}
\end{figure}

Figure \ref{fig2b} illustrates the various steps involved in the pre and post-processing of the hologram recorded using digital in-line holography. The processing method consists of three major steps:  (i) holographic reconstruction, (ii) network training, and (iii) post-processing of the holograms. As a first step, the background image is subtracted from the recorded holograms (see figure \ref{fig3}), followed by an intensity normalization process to remove noise and correct uneven illumination. The intensity normalization is represented as
\begin{equation}
I_{N}(x,y)=\frac{I-I_{min}}{I_{max}-I_{min}}
\end{equation}
where $I_{N}$ is the normalized image, and $I_{max}$ and $I_{min}$ are the maximum and the minimum intensities of the image. A set of 30  holograms without the particle field are recorded before the start of every experiment, which is then averaged to get the mean intensity image of the background. Once the holograms are pre-processed, to obtain the 3D optical field, the numerical reconstruction is performed using the Rayleigh-Sommerfeld equation, which is given by
\begin{equation}
I_{r}(x,y,z)=I_{h}(y,z)\otimes h(x,y,z),
\end{equation}
where, $I_{r}(x,y,z)$ is the 3D complex optical field obtained from reconstruction. The term $I_{h}(y,z)$ represents the pre-processed hologram, $\otimes$ denotes the convolution operation and $h(x,y,z)$ is the diffraction kernel. The Rayleigh-Sommerfeld diffraction kernel in the frequency domain can be expressed as \citep{katz2010applications}:
\begin{equation}
H(f_{x},f_{y},z)=exp\left (jkz\sqrt{1-\lambda^{2}f_{x}^{2}-\lambda^{2}f_{y}^{2}}\right)
\end{equation}
wherein, $j=\sqrt{-1}$ and $k=2\pi/\lambda$ represents the wavenumber; $\lambda $ is the wavelength of the incident beam. The spatial frequency in the $x$ and $y$ directions are expressed by $f_{x}$ and $f_{y}$, respectively. In the frequency domain, using the convolution theorem, the complex optical field (3D image of objects), $I_{r}$ can be evaluated as
\begin{equation}
I_{r}(x,y,z)={\rm FFT}^{-1} \left \{{\rm FFT} \left [I_{h}(y,z)\right]\times H(f_{x},f_{y},z)\right \}.
\end{equation}
Here, ${\rm FFT}$ represents the Fast Fourier Transform operator. The in-plane intensity information is obtained from the magnitude of the complex optical field in the corresponding plane. The reconstruction is performed in a series of planes with spacing $\Delta x=100~\mu$m across the test volume centered at the nozzle axis. The dimensions of the reconstructed volume are $45$ mm along the $x$ direction (along the optical axis), $40$ mm along the $y$ direction, and $31.3$ mm along the $z$ direction. The spatial resolutions for the shadowgraphy and holography images are 29.88 $\mu$m/pixels and 19.2 $\mu$m/pixels, respectively. They are calibrated using a calibration target (TSI Inc).

Figure \ref{fig3} shows a typical bag fragmentation of a water droplet under a straight airstream without swirl $(Sw=0)$ for $\We=14.53$. In this case, the dispensing needle is located at $(x_d/D_o,y_d/D_o,z_d/D_o) = (0.0, 0.27, 0.95)$. The temporal evolution of the fragmentation phenomenon obtained from shadowgraphy is shown in the top left panel of figure \ref{fig3}. It can be seen that the droplet enters the potential core region of the straight airstream at $\tau=-2.58$ and bulges to develop a bag of maximum size at the onset of its breakup at $\tau=0$. At $\tau=0.67$, bag and rim fragment into satellite droplets. At this instant, we illustrate the volume reconstruction, including the sample hologram and the reconstructed images at two different depths along the $x$ direction, as shown in figure \ref{fig3}. The insets in the reconstructed images at $x/D_0=0.9$ and -0.63 in figure \ref{fig3} highlight the in-focus images of the satellite droplets at different depths.

The second step (see figure~\ref{fig2b}) in holographic processing consists of training a  convolution neural network for segmenting the droplet images. A number of methods have been developed in the literature to determine the position and shape of droplets within the 3D reconstructed volume. For example, the droplet position and morphology are evaluated based on maximum edge sharpness and maximum intensity minimization \citep{guildenbecher2013digital,gao2013uncertainty,gao2013quantitative}. The common practice is to use edge detection followed by intensity thresholding to demarcate droplet boundaries in the images. Due to the unwanted noise surrounding the droplets region, conventional image processing techniques based on intensity gradients cannot fully segment the droplets. Additionally, if the concentration of droplets is dense, segmenting them based on traditional image processing techniques becomes difficult. To overcome the above-mentioned difficulties and to accurately discern the boundaries of droplets, the images are segmented using a 2D convolutional neural network based on the U-Net architecture \citep{Ronneberger2015,Falk2019}. The U-Net is a semantic segmentation method that was initially developed for analysing biological images \citep{Falk2019, Ibtehaz2020}. As part of its implementation, U-Net employs two paths for pixel-level classification and localization. By employing a series of convolutional and max-pooling layers, the first path captures image context, and the second path is an expanding path for detecting precise localization. Additionally, the data augmentation is implemented by elastically deforming the annotated input images. This allows the network to make better use of existing annotated data. An extensive discussion on the network architecture and its functionality can be found in the seminal work by \cite{Ronneberger2015}. In the present study, a total of 50 manually annotated (ground truth) images from the reconstructed volume are used to train the network. The ground truth annotation are performed using local thresholding around each satellite droplet. The set of manually annotated images from various cases enables the use of a single set of training weights for all experiments. Prior to the training, the current method \citep{Falk2019} performs data augmentation by rigidly and elastically deforming the ground truth images in order to eliminate the need for large data sets. The manual annotations or masks are created based on edge sharpness maximization. As annotating the entire image manually is difficult, only subzones of images are manually labeled and utilized for training. A GPU-based open-source software developed by \cite{Ronneberger2015} is utilized for training the network. On an NVIDIA P2200 GPU,  training the network takes approximately 12 hours.

The final step involves the post-processing of the 3D reconstructed volume in order to determine the droplet boundaries. The trained network is applied to every plane of the reconstructed 3D volume as shown in figure~\ref{fig2b}. The network's output directly provides the binary masks corresponding to the droplet boundaries in each plane. The final processing involves the maximum intensity projection of the binary mask, elimination of droplets smaller than 3 pixels \citep{berg2022tutorial}, and estimation of the equivalent diameter. \ks{The network validation using synthetic holograms and the associated errors are provided in the supplementary information (Figure S3).}

\begin{table}
\begin{center}
\begin{tabular}{cccc}
 Fluids & Density, $\rho$ (kg/m$^3$) & Viscosity, $\mu$ (mPa$\cdot$s) & Surface tension, $\sigma$ (mN/m) \\ 
 Water           & 998           & 1.0           & 72.8                \\  
 Ethanol     & 789        & 1.2  & 24.4       \\  
\end{tabular}
\end{center} 
\caption{The fluid properties of water and ethanol used in the present study and considered by \cite{guildenbecher2017characterization}, respectively.} \label{T1}
\end{table}

\begin{figure}
\centering
\includegraphics[width=0.7\textwidth]{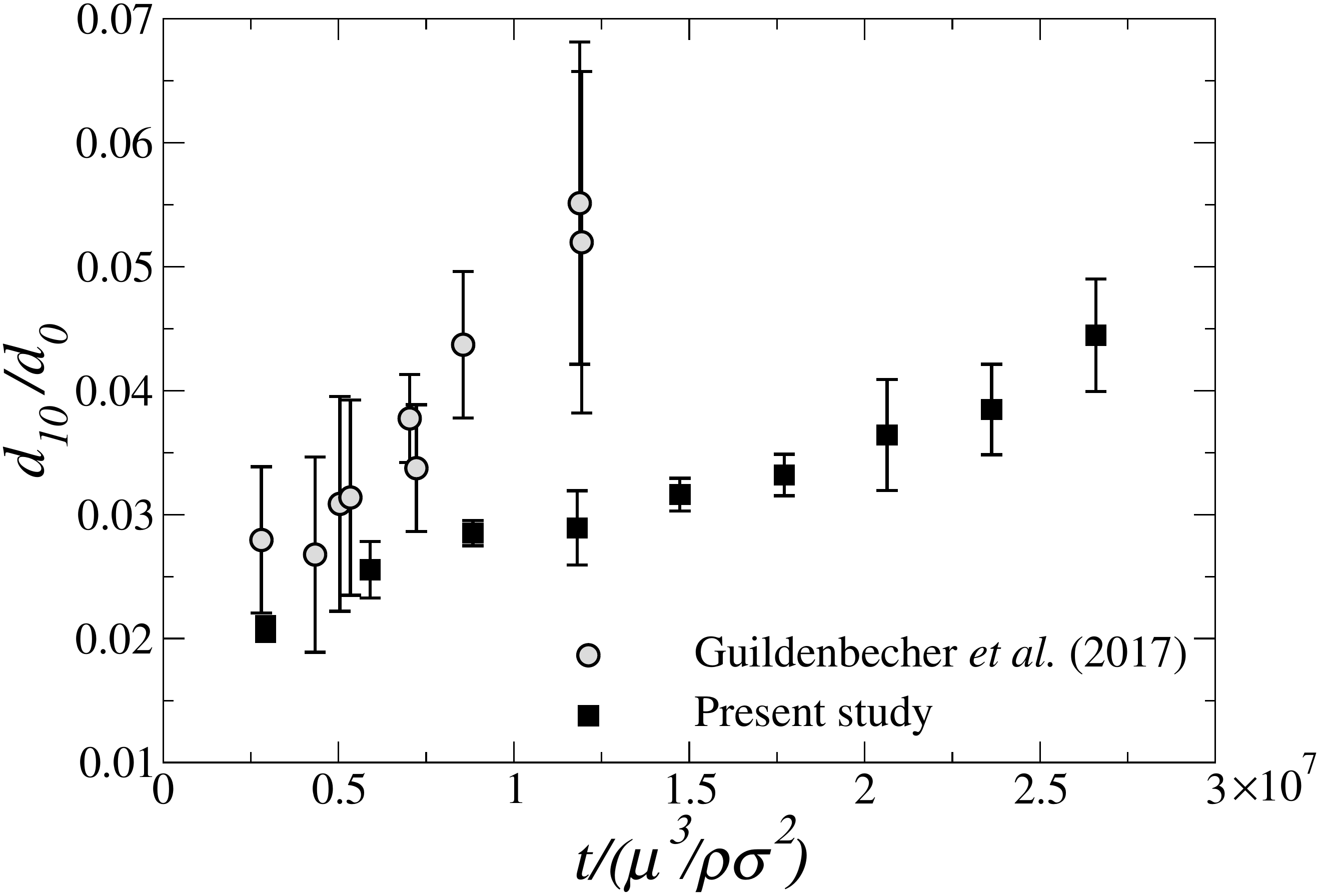}
\caption{Comparison of the temporal variation of normalised mean diameter $(d_{10}/d_0)$ obtained from our experiments with that of \cite{guildenbecher2017characterization}. In the present study, a water droplet of $d_0 = 3.09$ mm ($\We = 14.53$) is used, while, in \cite{guildenbecher2017characterization}, an ethanol droplet of $d_0 = 2.54$ mm ($\We = 13.8$) was considered. In our experiments, the error bar represents the standard deviation obtained from the three repetitions. \ks{The uncertainty bar in \cite{guildenbecher2017characterization} corresponds to the standard deviation from forty-four realisations.}}
\label{fig4}
\end{figure}

In order to validate the results obtained from our analysis, we compare the size distribution of satellite droplets resulting from the fragmentation of a water droplet under a straight airstream (without swirl) in a cross-flow configuration with those reported in \cite{guildenbecher2017characterization}. It is to be noted that we consider a water droplet of $d_0 =3.09$ mm ($\We = 14.53$) while \cite{guildenbecher2017characterization} used an ethanol droplet of $d_0=2.54$ mm ($\We = 13.8$). The properties of both water and ethanol are listed in Table \ref{T1}. Figure \ref{fig4} depicts the comparison of the variation of the normalized mean diameter ($d_{10}/d_0$) with normalised time $(t/(\mu ^{3}/\sigma ^{2}\rho))$. In this figure, we have adopted a normalisation for time based on fluid properties to compare the results associated with different working fluids considered in our experiments (water) and \cite{guildenbecher2017characterization} (ethanol). Here, $d_{10}$ is a number-based spatial average of the diameters of all droplets and is given by
\begin{equation} \label{R:eq9}
d_{10}=\int_{0}^{\infty}dp(d)\textrm{d}d,
\end{equation} 
where $p(d)$ is the probability density function of the diameters of the satellite droplets. \ks{We calculate $d_{10}$ from the number-based-average of satellite droplets.} It can be seen in figure \ref{fig4} that at early times the ratio $d_{10}/d_0$ is small due to the rapture of the bag near its tip. The fragmentation of the rim and nodes occurs in later stages, providing several larger satellite droplets. This results in all the droplets contributing to the estimation of $d_{10}$, increasing the normalized mean diameter of the satellite droplets, $d_{10}/d_0$. It can also be observed that at initial times, the estimated $d_{10}/d_0$ from the present experiment is in reasonable agreement with that of \cite{guildenbecher2017characterization}. However, the results differ at the later stages of fragmentation. It is to be noted that the total breakup time of the water droplet in our experiments compares well with \cite{kulkarni2014bag}. Quantitatively, the total breakup time obtained in our experiment is 4.44 ms, while it is 6 ms in the experiment of \cite{kulkarni2014bag}. The Rayleigh-Taylor instability develops as a result of the penetration of the air phase into the liquid phase, which causes the fragmentation of the liquid droplet. The Atwood number, $At={(\rho-\rho_a)/(\rho+\rho_a)}$ for the water-air and ethanol-air systems are 1.2 and 0.997, respectively. As a result, one would expect a more substantial Rayleigh-Taylor instability to be developed in the case of the water-air system, as considered in our study, than the ethanol-air system considered by \cite{guildenbecher2017characterization}. This, in turn, helps a water droplet fragment into smaller satellite droplets more easily than an ethanol droplet. Thus, it can be observed in figure \ref{fig4} that the normalized mean diameter of the satellite droplets in our experiment (water droplet) is smaller than that reported by \cite{guildenbecher2017characterization} for an ethanol droplet. The difference in fluid properties of the liquids considered in our experiment and in \cite{guildenbecher2017characterization} may not significantly affect the rapture of the thin film of the bag but mainly influences the fragmentation of the rim breakup phenomenon. While the rapture of the bag is due to the pressure difference across its interface caused by the aerodynamic force, the later stage of rim fragmentation is driven by the capillary Rayleigh–Plateau instability \citep{jackiw2021aerodynamic}. The values of the Ohnesorge number, $Oh=\mu/\sqrt{\rho \sigma d_0}$ in our case and in the study of  \cite{guildenbecher2017characterization} are about 0.002 and 0.005, respectively. Therefore, at the later stage, the rim fragmentation in our study is slower than that reported in the work of \cite{guildenbecher2017characterization}.  

\section{Results and discussion}
\label{sec:dis}
\begin{figure}
\centering
\includegraphics[width=0.8\textwidth]{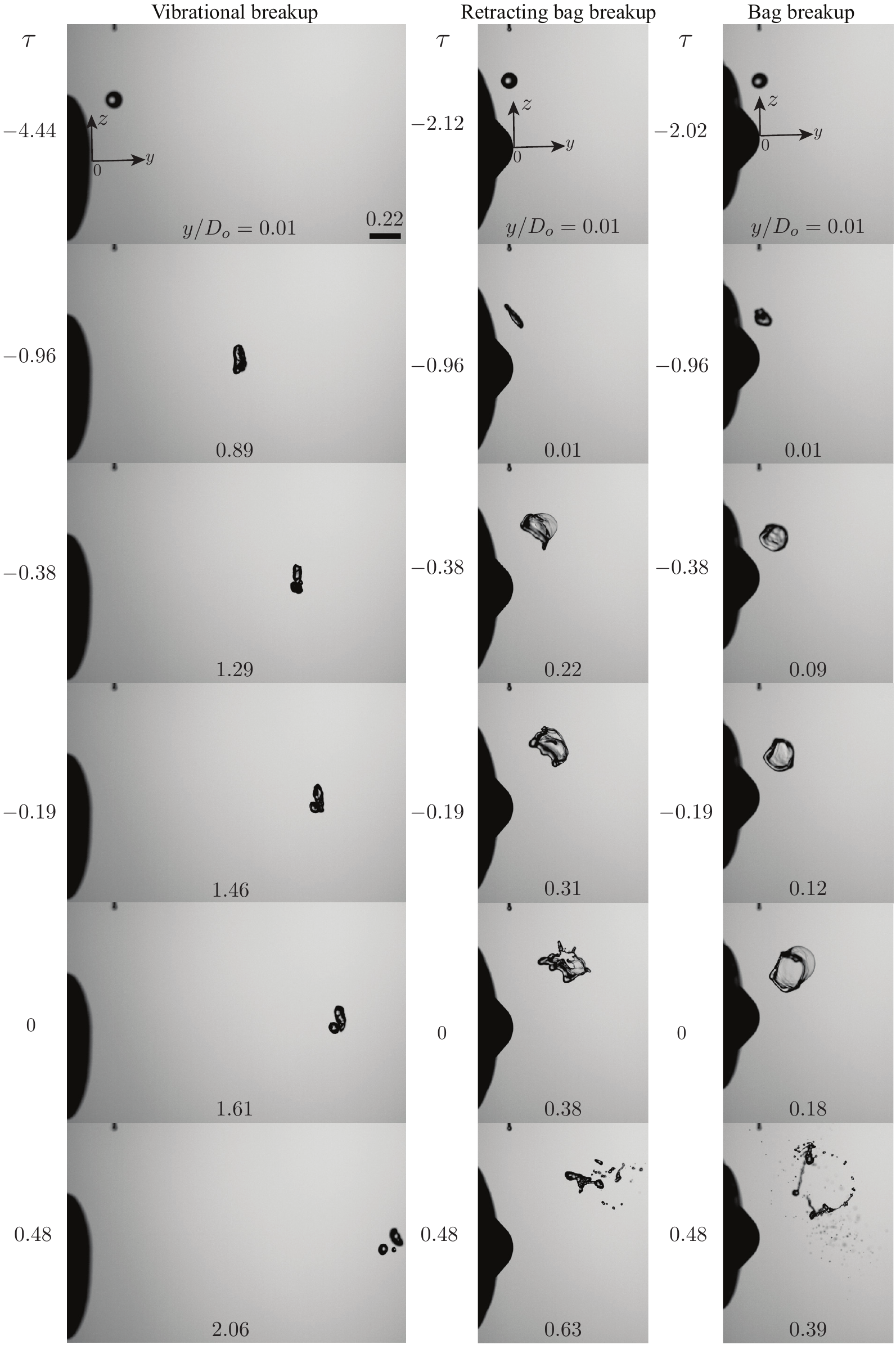}
\caption{Effect of the Swirl number, $Sw$ on the fragmentation process of a water droplet (obtained from shadowgraphy) for $\We = 12.1$. First column: $Sw = 0$ (vibrational breakup), second column: $Sw = 0.47$ (retracting bag breakup) and third column: $Sw = 0.82$ (normal bag breakup). The value of $\tau$ is indicated in each panel. The dimensionless scale-bar is shown on the top-left panel. The dimensionless location $(x_d/D_o,y_d/D_o,z_d/D_o)$ of the dispensing needle for $Sw=0$, $0.47$ and $0.82$ are $(0.0,0.01,0.80)$, $(0.12,0.01,0.80)$ and $(0.12,0.01,0.80)$, respectively. To show the migration due to the airstream, the dimensionless position of the droplet in the flow direction, $y/D_0$ is mentioned at the bottom of each panel. \cite{kirar2022experimental} have shown similar breakup modes for an ethanol droplet. The droplet breakup phenomena for $Sw=0$, 0.47 and 0.82 are provided as supplementary movies 2, 3 and 4, respectively.}
\label{fig2}
\end{figure}

As discussed in the introduction, \cite{kirar2022experimental} employed shadowgraphy to illustrate different breakup modes for an ethanol droplet of diameter $d_0=2.7 \pm 0.07$ mm and discovered a new breakup mode (retracting bag breakup mode) for a low value of the Swirl number $(Sw)$. To the best of our knowledge, the study by  \cite{kirar2022experimental} is the first to explore the phenomenon of droplet breakup in a swirling airstream. In contrast to the present study, which focuses on droplet size distribution, their aim was to demonstrate a new breakup mode in a swirling flow using Raleigh-Taylor instability. Thus, before discussing the size distribution of droplets obtained using the digital in-line holography technique, we begin the presentation of our results by demonstrating the temporal evolution of the morphologies of a freely falling water droplet of diameter $d_0=3.09 \pm 0.07$ mm interacting with an airstream of different Swirl strengths. Figure \ref{fig2} depicts the temporal evolution of the droplet morphology for the no-swirl ($Sw=0$), low swirl ($Sw=0.47$), and high swirl ($Sw=0.82$) conditions for a fixed value of the Weber number ($\We = 12.1$). The dimensionless locations of the tip of the dispensing needle $(x_d/D_o,y_d/D_o,z_d/D_o)$ are $(0.0,0.01,0.80)$, $(0.12,0.01,0.80)$ and $(0.12,0.01,0.80)$ for $Sw=0$, $0.47$ and $0.82$, respectively. In the no-swirl case, the droplet is dispensed at the axis of symmetry, $x_d/D_o=0$, but in swirl flow cases, the droplet is dispensed at $x_d/D_o = 0.12$, so that the droplet interacts with the swirl airstream in oppose/cross-flow conditions. The location of the dispensing needle for the no-swirl and swirl cases is shown in figure S2 of the supplementary information. These locations are chosen in accordance with the regime map presented in \cite{kirar2022experimental} (see, \ks{figure S4} in the supplementary material). The dimensionless time, $\tau \left (\equiv  t /t_d \right)$ is mentioned at the top of each panel in figure \ref{fig2}. Here $t$ is the physical time and $t_{d}=d_{0}\sqrt{\rho /\rho _{a}}/U$ is the characteristic deformation time. In our experiments, $\tau=0$ represents the onset of the breakup of the droplet. In the case of bag breakup mode, it is the instant the inflated bag ruptures, but in the case of vibrational breakup mode, it is the instant the droplet starts to break into smaller droplets of comparable size. Thus, the value of $\tau$ before the fragmentation instant is a negative number.

\begin{figure}
\centering
\includegraphics[width=0.6\textwidth]{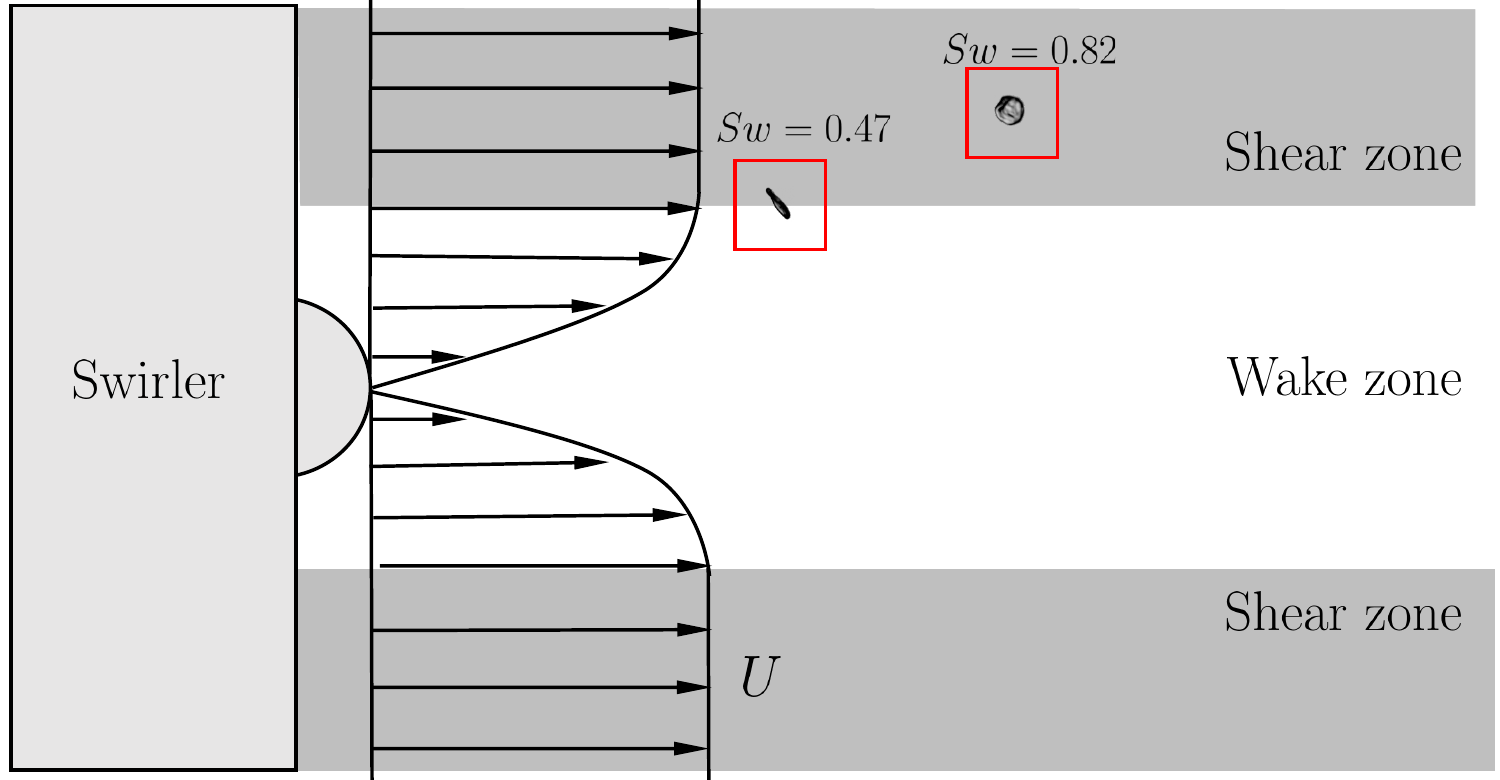}
\caption{Schematic of the core-annular flow field of the high shear and wake zones created due to the swirling effect. The droplets during its interaction with the swirl airstream for low and high swirl strengths are also depicted.}
\label{fig2a}
\end{figure}

It can be seen in figure 5 that the droplet undergoes a vibrational breakup mode in the no-swirl case (first row), but it exhibits a retracting bag breakup mode for $Sw=0.47$ (middle row) and a normal bag breakup mode for $Sw=0.82$ (bottom row). In the no-swirl case, as the spherical droplet enters the aerodynamic field, it deforms into a thick disk ($\tau=-0.96$) as a result of asymmetrical pressure distribution at the front and rear of the droplet. At this stage, the surface tension force opposes further deformation and try to bring the droplet to a spherical shape. Consequently, the droplet oscillates with an increasing amplitude due to the competition between the aerodynamic force favouring the deformation, and the viscous and surface tension forces opposing it (at $\tau =-0.38$ and $-0.19$). As the oscillations reach an amplitude comparable to the drop's radius, it fragments into smaller satellite droplets.

For the low swirl strength ($Sw=0.47$), the droplet quickly changes its shape from a sphere to a slightly tilted disk due to the aerodynamic force in the direction of vane inclination. As the swirl is not strong enough, only the bottom part of the disk enters the low-velocity zone (schematically shown in figure \ref{fig2a}) generated by the wake of the swirler. Consequently, the rest of the disk, which is in the high-shear region, undergoes inflation to form a bag. In the high shear zone, a negative pressure gets created that retracts the bag sheet in a direction opposite to the direction of bag growth ($\tau =-0.19$). Subsequent retraction of the bag exhibits capillary instability and causes the bag to rupture ($\tau=0$). When the bag ruptures, the surface tension pulls the liquid film surrounding it to the rim, causing it to break into satellite droplets ($\tau=0.48$).

For the high swirl strength ($Sw=0.82$), the droplet rapidly changes its topology from a sphere to a disk shape due to the unequal pressure distribution around its periphery ($\tau =-2.02$ to $-0.96$). The lower portion of the disk tilts as it is exposed to an opposed flow condition. In this case, the swirling strength is sufficient to retain the disk in the shear zone. In other words, the high swirl flow does not allow the disk to enter into the low-velocity zone created by the wake of the swirler (see figure \ref{fig2a}). As a result, the entire disk adjusts vertically, making the breakup process more efficient ($\tau =-0.38$ to $-0.19$). The middle of the disk elongates, generating a thin liquid sheet in the centre and a thick rim on the outside ($\tau =0$). Finally, high shear velocity causes the bag and rim to break in the shear zone ($\tau =0.48$).

\begin{figure}
\centering
\includegraphics[width=0.95\textwidth]{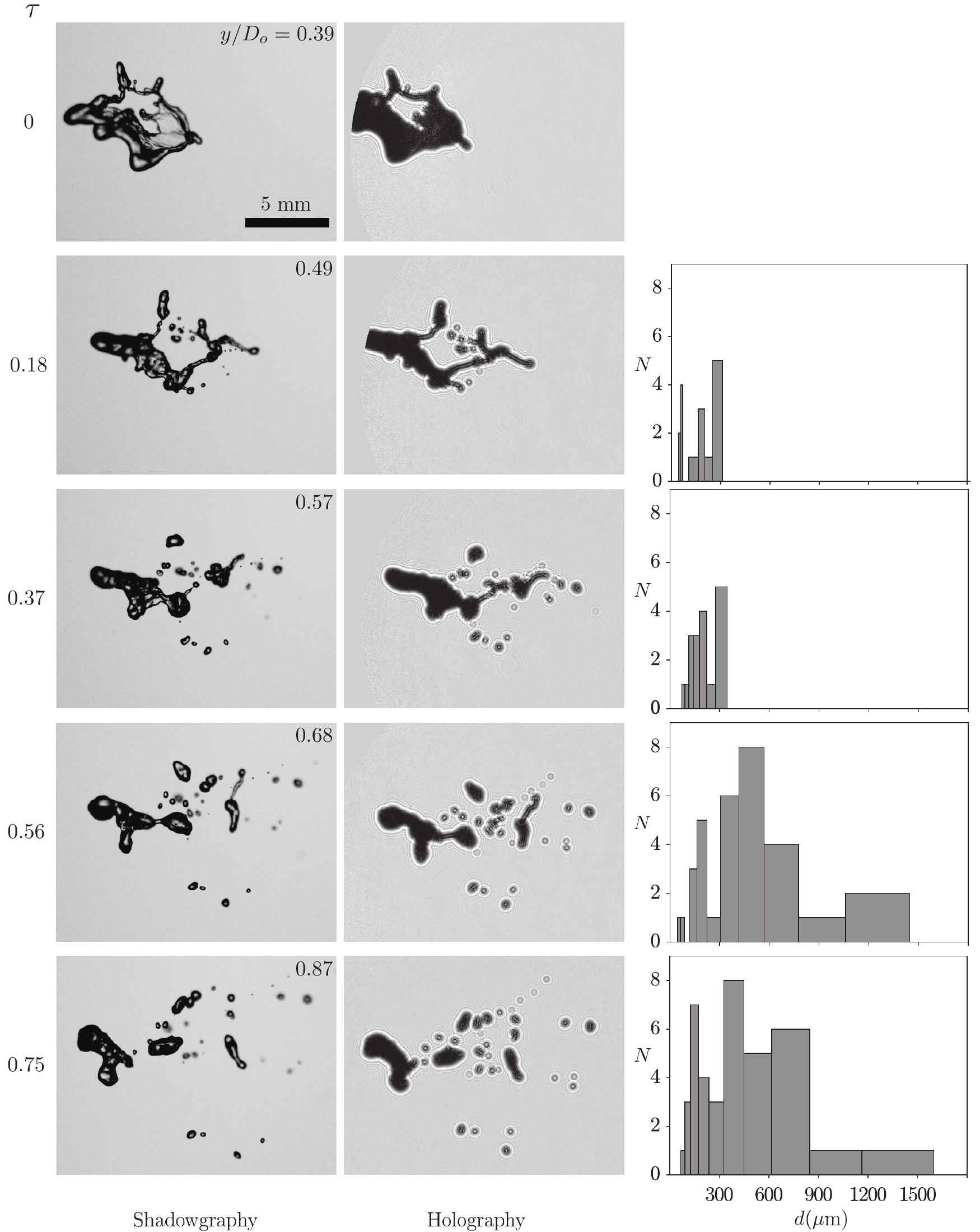}
\caption{Temporal evolution of the droplet size distribution for $Sw = 0.47$ and $\We = 12.1$. The first and second columns represent the shadowgraphy images and recorded holograms obtained using the in-line holography. The values of the dimensionless time, $\tau$, are mentioned on the left side of the first column measured from the instant at the onset of breakup. The scale-bar is shown in the top-left panel. The third column depicts the histograms of the droplet size distribution (the droplet counts, $N$ versus the droplet diameter, $d$) at different instants after the breakup. The dimensionless location of the dispensing needle is at  $(x_d/D_o,y_d/D_o,z_d/D_o) = (0.12, 0.01, 0.80)$. The dimensionless position of the droplet in the flow direction, $y/D_0$ is mentioned at the top-right corner of each panel of the shadowgraphy images.}
\label{fig5}
\end{figure}

Next, we investigate the size distribution of a droplet after fragmentation under different swirl air strengths. Figure \ref{fig2} demonstrates that the droplet disintegrates into a few large satellite droplets in the no-swirl situation for $\We=12.1$ due to the vibrational breakup. Therefore, in the following, we mainly focus on the size distribution in the low and high swirl strengths that encounter retracting bag and normal bag breaking events. Figure \ref{fig5} depicts the temporal evolution of the droplet size distribution for $Sw = 0.47$  (low swirl strength) at $\We = 12.1$. In figure \ref{fig5}, the first, second and third columns show the shadowgraphy images, holograms, and the resultant size distributions of droplets at different instants, respectively. At $\tau=0$, the retracted bag ruptures due to the negative pressure gradient created by the wake of the swirler. It is interesting to note that, although the bag is ruptured, it does not produce any satellite drops at this stage ($\tau =0$). This phenomenon is distinct from the normal bag breakup observed in a straight airstream at high Weber numbers (see, for instance, \cite{guildenbecher2009secondary,kulkarni2014bag,soni2020deformation}). For $\tau > 0$, it can be seen in figure \ref{fig5} that the bag sheet continues to retract and finally impinges on the rim. This leads to the fragmentation of the upper portion of the rim due to the capillary Rayleigh–Plateau instability. The fragmentation of the rim continues which leads to the generation of smaller secondary droplets of diameter upto 300 $\mu$m, as indicated by the histogram at $\tau =0.37$ in figure \ref{fig5}. During this process, the nodes formed on the rim due to the Rayleigh-Taylor instability also disintegrates, and this process continues till $\tau =0.56$ thereby producing more droplets in broader size range, $300 ~\mu\textrm{m} < d < 1200 ~\mu\textrm{m}$. Finally, as the bottom of the disk is entrapped in the wake zone, it creates a low aerodynamic force that leads to thicker nodes. These thicker nodes (at $\tau =0.75$) break eventually because of the continuous swirl airflow, which in turn, creates larger satellite droplets ($d>1200$ $\mu\textrm{m}$). Note that the histogram is truncated at $d \approx 1600$ $\mu\textrm{m}$, and the distribution of larger droplets is not shown here. A volume-weighted distribution presented later in figure \ref{fig9} provides the entire size distribution of the droplets.

\begin{figure}
\centering
\includegraphics[width=0.95\textwidth]{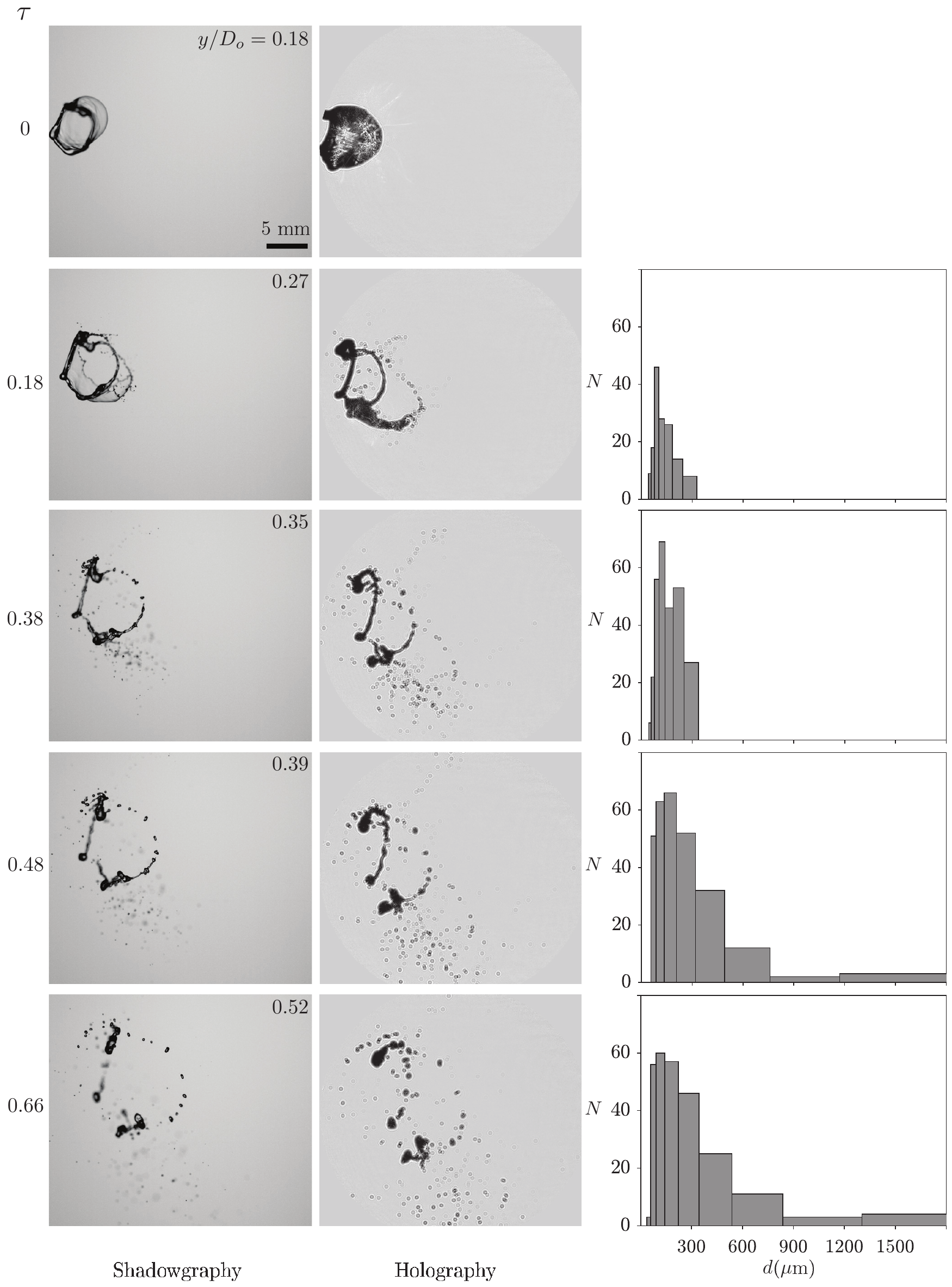}
\caption{Temporal evolution of the droplet size distribution for $Sw = 0.82$ and $\We = 12.1$. The first and second column represent the shadowgraphy and the holography (recorded hologram) images, respectively. The values of the dimensionless time, $\tau$ are mentioned on the left side of the first column measured from the instant at the onset of breakup. The scale-bar is shown on the top-left panel. The third column depicts the histogram of the droplet size distribution (the droplet counts, $N$ versus the droplet diameter, $d$) at different instants after breakup. The dimensionless location of the dispensing needle is $(x_d/D_o,y_d/D_o,z_d/D_o) = (0.12, 0.01, 0.80)$. The dimensionless position of the droplet in the flow direction, $y/D_0$ is mentioned at the top-right corner of each panel of the shadowgraphy images.}
\label{fig6}
\end{figure}

Figure \ref{fig6} illustrates the temporal evolution of the droplet size distribution for $Sw = 0.82$ (high swirl strength) for the same Weber number ($\We = 12.1$). In this case, the entire liquid bulk exposes to the shear zone (see the droplet position in figure \ref{fig2a}) and changes its morphology from a disk to a thick toroidal rim with an elongated bag ($\tau =0$). The bursting of this bag ($\tau =0.18$) generates very fine satellite droplets $(d<300$ $\mu\textrm{m})$. Subsequently, the remaining portion of the bag liquid sheet propagates back towards the rim due to unbalanced tension and is finally absorbed in the rim. At this stage ($\tau =0.38$), the number of tiny satellite droplet increases, and fragmentation of rim initiates. The breakup of the rim advances and results in generation of intermediate-sized droplets ($300 ~ \mu\textrm{m} <d<900 ~ \mu\textrm{m}$) at $\tau=0.48$. Finally, due to the Rayleigh-Taylor instability, the nodes on the rim disintegrate and produce bigger satellite droplets ($d>900 ~ \mu\textrm{m}$) at $\tau =0.66$. At this stage, the coalescence of smaller droplets also contributes to the size distribution.
 
\begin{figure}
\centering
\hspace{0.2cm} {\large (a)} \hspace{5.9cm} {\large (b)} \\
\includegraphics[height=0.32\textwidth]{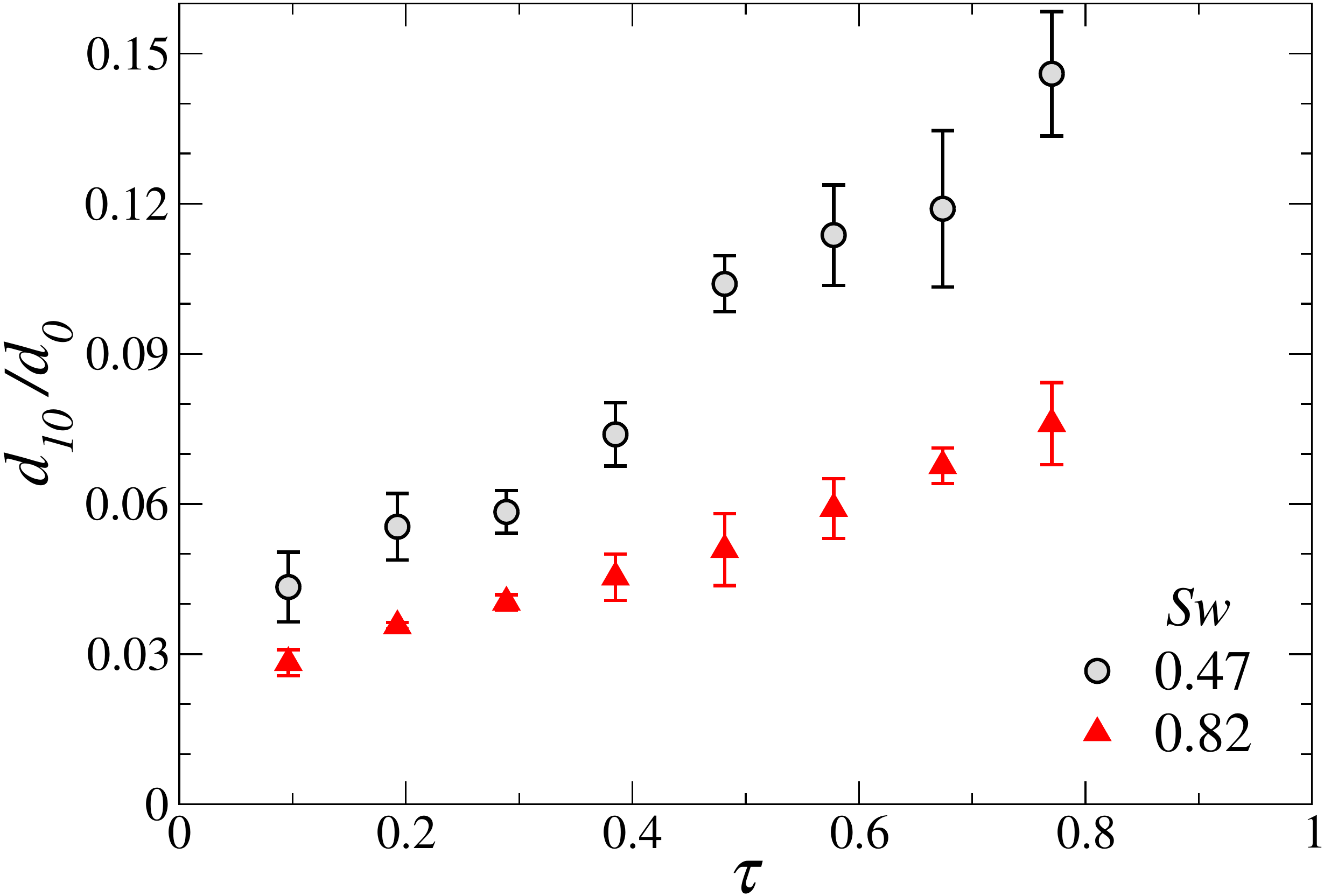}  \includegraphics[height=0.32\textwidth]{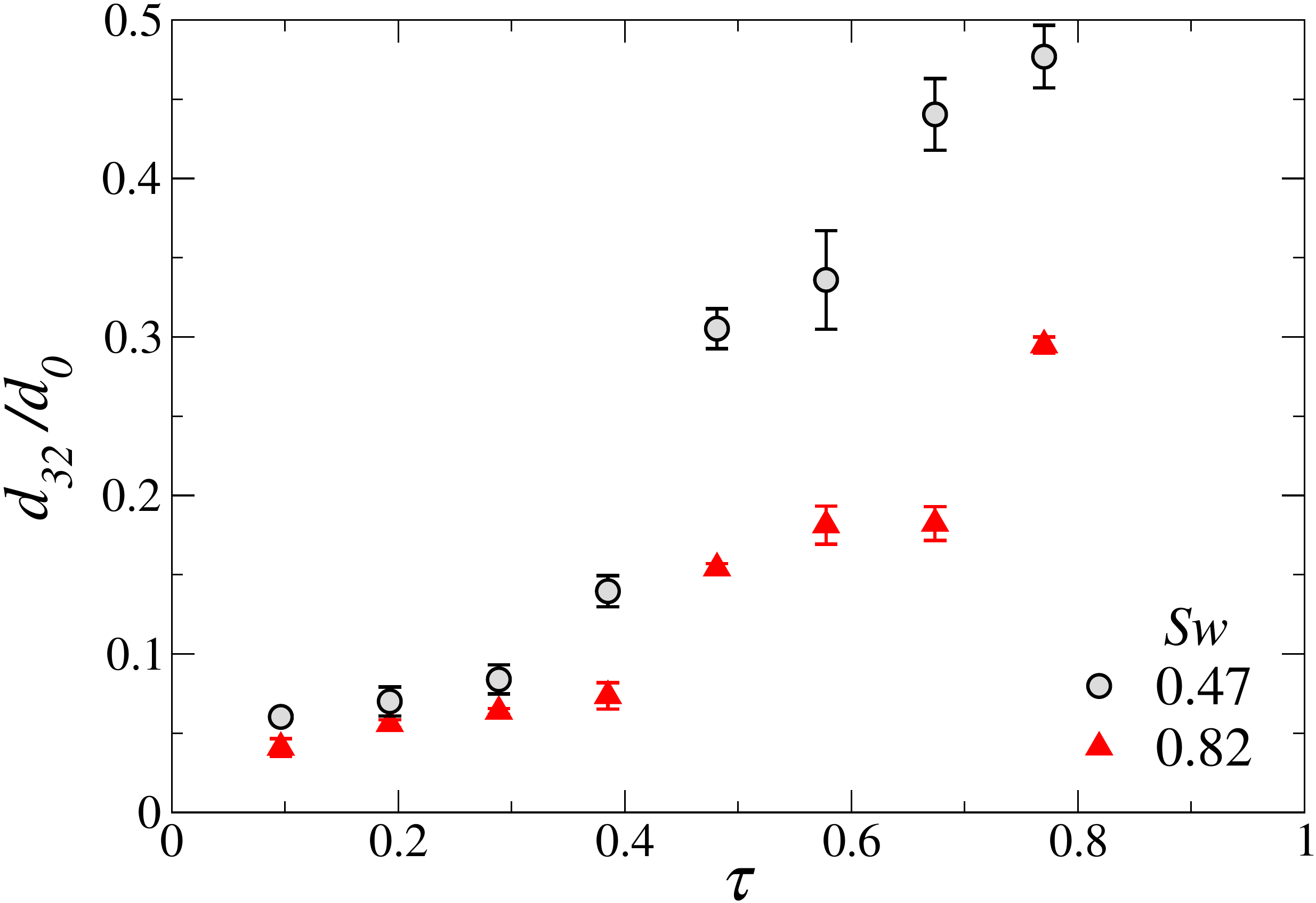}
\caption{Temporal variation of normalised mean diameters (a) $(d_{10}/d_0)$ and (b) $(d_{32}/d_0)$ for different swirl numbers at $\We = 12.1$.}
\label{fig7}
\end{figure}

Figure \ref{fig7}(a) shows temporal variation of normalised number-mean diameter for different Swirl numbers at $\We=12.1$. The error bar at each data point represents the standard deviation of three repetitions. As time progresses, the relative velocity between airflow and liquid bulk decreases, thus producing bigger droplets leading to an increase in the number-mean diameter of satellite droplets. As discussed earlier, while the low swirl case ($Sw=0.47$) produces larger size droplets due to the fragmentation of rim and nodes (bag-film breakup does not contribute to the size distribution), in the high swirl case ($Sw=0.82$), the breakup of the bag, rim, and nodes contribute to the number-mean diameter and thereby producing smaller satellite droplets. 

The Sauter mean diameter (commonly termed as SMD or $d_{32}$) is defined as  surface area moment mean, which provides the mean size of a given droplet  distribution. Thus, the reciprocal of the Sauter mean diameter is a direct measure of the surface area per unit volume of the satellite droplets, which is given by 
\begin{equation} \label{Sauter}
d_{32}=\frac{\int_{0}^{\infty}d^{3}p(d)\textrm{d}d}{\int_{0}^{\infty }d^{2}p(d)\textrm{d}d}.
\end{equation}
\ks{In the present study, we use the ratio between the mean volume and mean surface area of the satellite droplets to calculate $d_{32}$.} 

To get a better insight into the size distribution of satellite droplets, we plot the temporal variation of normalised Sauter mean diameter ($d_{32}/d_{0}$) for different swirl numbers in figure \ref{fig7}(b). It can be seen that the Sauter mean diameter increases with time for both the low and high swirl strengths. However, in comparison to low swirl strength, high swirl strength produces secondary droplets with a smaller value of $d_{32}$. Therefore, for the same aerodynamic force, the high swirl strength generates more surface area (and hence more surface energy) than the low swirl strength after fragmentation. At a later time ($\tau =0.5$), the sudden jumps in the variation of $d_{32}/d_{0}$ is associated with the increase in the number of larger droplets.   

\subsection{Prediction of droplet size distribution from statistical theory}
In this section, we compare our experimentally determined droplet size distribution in swirl airstreams with an existing theoretical model, which was initially developed for the no-swirl conditions \citep{Villermaux2009single}. According to \cite{Villermaux2009single}, the corrugated rims contribute to the overall size distribution, regardless of the mode of breakup (whether it is a bag breakup or a direct transition from drop to ligament). In a strong shear flow, the detached ligament from the corrugated rim elongates and disintegrates into numerous blobs due to capillary instability. \cite{marmottant2004spray} remarked that during this phase, the blobs of various shapes and sizes also coalesce, leading to the formation of larger droplets because of the Laplace pressure difference. Therefore, the resultant droplets have a diameter greater than the thickness of the ligament. All these processes are known to affect the overall size distribution that follows a single parameter gamma distribution function \citep{villermaux2007fragmentation} as
\begin{equation} \label{v:eq3}
P_{b}=\frac{\nu ^{\nu }}{\Gamma (\nu )}\left ( \frac{d}{d_{1}} \right )^{\nu -1}e^{-\nu (d/d_{1})},
\end{equation} 
where $\nu=1/(\gamma -1)$ reflects the regularity of the initial shape of the ligament, wherein $\gamma$ is the interaction parameter and $d_1$ is the average blob diameter. For a corrugated ligament, $\gamma> 1$ that induces a skewed broader distribution of sizes with an exponential tail. Its value approaches to 1 as the ligament is smoother. Thus the value of $\nu$ increases with the smoothness of the ligament, which in turn give rise to a uniform distribution of drop sizes. The gamma distribution closely fits with the present experiments when the value of $\gamma=1.25$. The larger droplets are distributed exponentially in the distribution shown above. There is a high likelihood that these droplets will keep disintegrating until they stabilise. Due to the consecutive breaking up of large droplets, the total size distribution that results from this follows an exponential distribution as \citep{Villermaux2009single}
\begin{equation} \label{v:eq4}
P(d)=\int P_{b}\frac{e^{-d_{1}/\left \langle d \right \rangle}}{\left \langle d \right \rangle}\textrm{d}d_{1},
\end{equation} 
where $P(d)$ is the number probability density. Thus, the dimensionless number probability density, $p(\chi)=\left \langle d \right \rangle P(d)$, is giving by
\begin{equation} \label{v:eq5}
p(\chi)=\frac{32}{3}\chi^{^{\frac{3}{2}}}K_{3}(4\sqrt{\chi}),
\end{equation} 
where $K_3 $ represents the Bessel function of third order, $\left \langle d \right \rangle$ denotes the average drop size and $\chi=d/\left \langle d \right \rangle$.

\begin{figure}
\centering
\hspace{0.4cm} {\large (a)} \hspace{5.8cm} {\large (b)} \\
\includegraphics[height=0.32\textwidth]{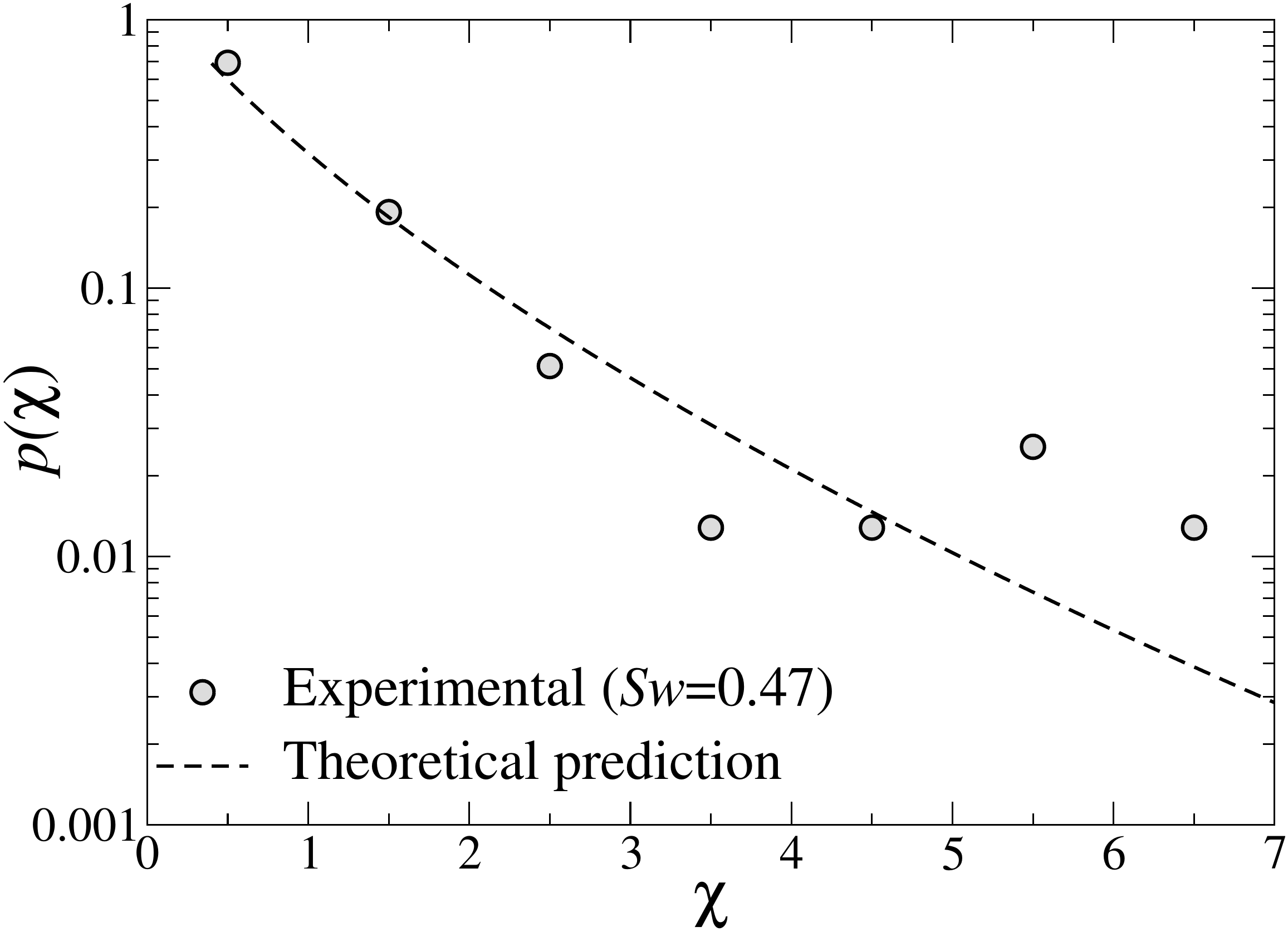} \hspace{2mm} \includegraphics[height=0.32\textwidth]{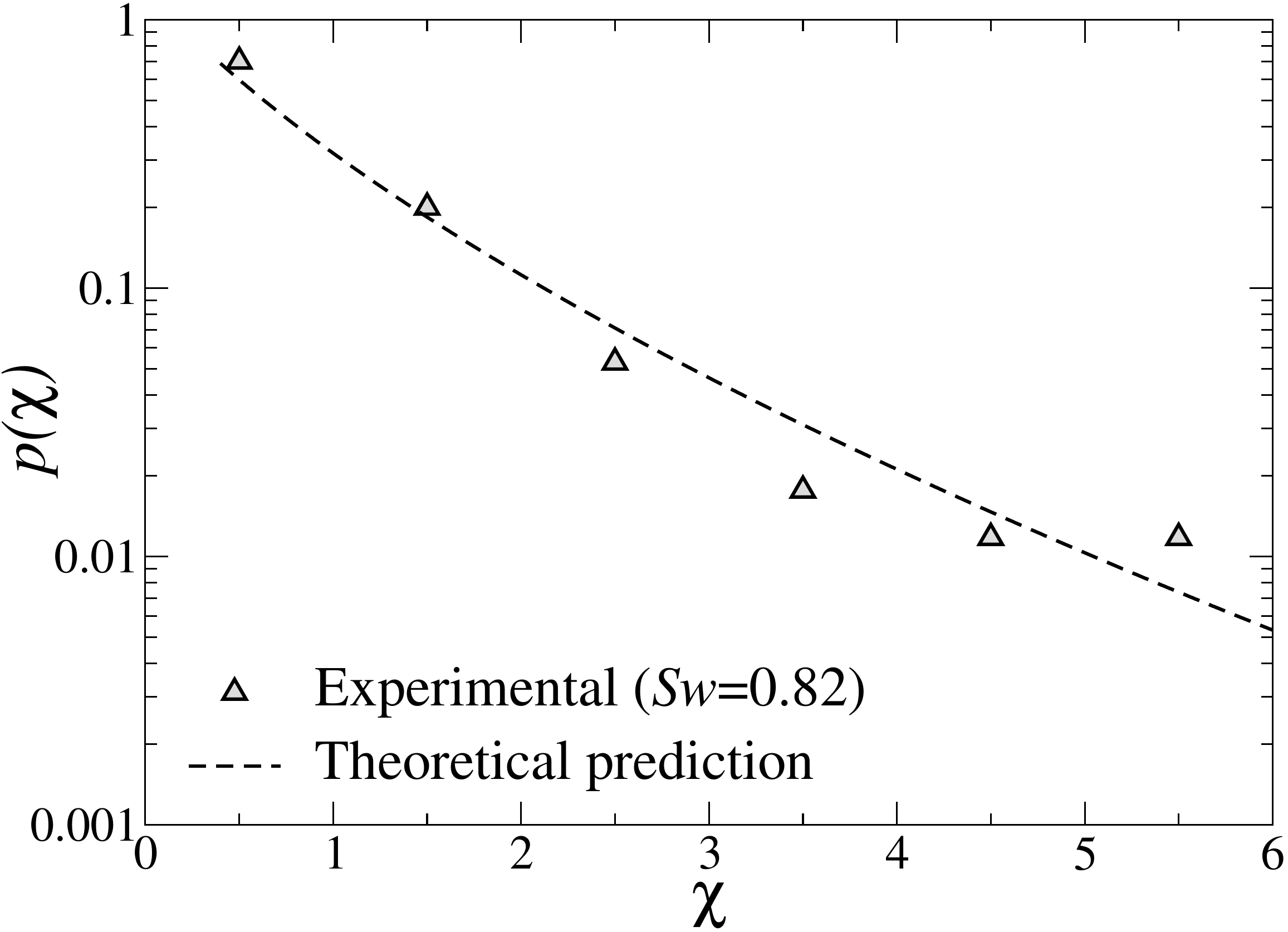}
\caption{Comparison of the variation of $p(\chi)$ versus $\chi\left (d/\langle d \rangle \right)$ obtained from our experiments with the theoretical prediction from Eq. (\ref{v:eq5}) \citep{Villermaux2009single} for (a) $Sw = 0.47$  at $\tau = 0.75$ and (b) $Sw = 0.82$ at $\tau = 0.66$. The rest of the parameters are the same as figure \ref{fig5}.}
\label{fig8}
\end{figure}

Figure \ref{fig8}(a) and \ref{fig8}(b) depicts the comparison of  dimensionless number probability density, $p(\chi)$ obtained from our experiments with the theoretical prediction (Eq. \ref{v:eq5}) for the low swirl ($Sw=0.47$) and high swirl ($Sw=0.82$) strengths, respectively. For both cases, a typical instant is chosen when the fragmentation ceases. It can be observed in figure \ref{fig8}(a) that for low swirl strength ($Sw=0.47$) that exhibits retracting bag type breakup, the theoretical prediction agrees with the experimental observation for small droplets ($\chi < 3$), but deviates from the experimental results for large droplets. As discussed earlier, in the case of retracting bag breakup, the lower portion of the disk/rim is entrapped in the wake zone of the swirler and does not participate in the bag formation process. This results in producing larger droplets and alters the size distribution as observed for $\chi > 3$ in figure \ref{fig8}(a). Additionally, the rim instability and subsequent droplet count may be impacted by the bag's retraction and impingement on the rim. This phenomenon is entirely missing in the event of a normal bag breakup. This is evident in figure \ref{fig8}(b) for $Sw=0.82$ that undergoes a normal bag breakup and shows a better agreement with the theoretical prediction. However, a close inspection reveals that the theoretical prediction differs slightly from the experimental results for very large droplets ($\chi > 3$). It is due to the fact that all breakup modes are not taken into consideration in theoretical prediction \citep{Villermaux2009single}.

\subsection{Multi-modal size distribution}
In the preceding section, we have analysed the nature of the size distribution based on the droplet counts by fitting a single parameter gamma distribution. However, the single parameter distributions cannot be used to distinguish between various mechanisms responsible for droplet breakup. Additionally, the number mean diameter is a required input parameter for this distribution. \cite{jackiw2022prediction} developed an analytical model to predict the combined multi-modal distribution, accounting for various mechanisms, for the aerodynamic breakup of droplets in a no-swirl condition.

In the present study, for swirl airstreams, we have followed a similar approach as that of  \cite{jackiw2022prediction} by considering the volume-weighted probability density function, $P_v$ instead of the number-weighted probability density function, $P_n$, as the distribution based on $P_n$ is more biased towards the smaller size droplets. The volume-weighted density refers to the ratio of the total volume of droplets of a specific diameter to the total volume of all droplets. Thus, $P_v$ is given by
\begin{equation} \label{j:eq1}
P_{v}=\frac{\zeta ^{3}P_{n}}{\int_{0}^{\infty}\zeta ^{3}P_{n}d\zeta}={\frac{\zeta ^{3}P_{n}}{\beta^{3}\Gamma (\alpha+3)/\Gamma (\alpha)}},
\end{equation}
where $\zeta \left (=d/d_0 \right)$ denotes the normalized droplet size, $\Gamma$ represents the gamma function, and  $\alpha=(\bar{\zeta}/\sigma_s)^{2}$ and $\beta=\sigma_s^{2}/\bar{\zeta}$ are the shape and rate parameters, respectively. Here, $\bar{\zeta}$ and $\sigma_s$ are the mean and standard deviation of the distribution. The expression for $P_n$ is given by
\begin{equation} \label{j:eq2}
P_n=\frac{\zeta^{\alpha -1}e^{-\zeta /\beta}}{\beta^{\alpha}\Gamma (\alpha)}.
\end{equation}
In the case of bag fragmentation, three modes, namely, the node, rim, and bag breakup modes, contribute to the overall size distribution of satellite droplets. Therefore, it is essential to take into account their relative contribution to the size distribution by considering the weighted summation of each mode. The total $P_v$ determined by the weighted sum of each mode is given by
\begin{equation} \label{j:eq3}
P_{v,Total}=w_{N}P_{v,N}+w_{R}P_{v,R}+w_{B}P_{v,B},
\end{equation}
where $w_{N}=V_{N}/V_{0}$, $w_{R}=V_{R}/V_{0}$ and $w_{B}=V_{B}/V_{0}$ represent the contributions of volume weights from the node, rim and bag, respectively. Here, $V_{N}$, $V_{R}$, $V_{B}$ and $V_{0}$ are the node, rim, bag and the initial droplet volumes, respectively. 

To begin with, we will elucidate the volume weight of each mode. The volume weight corresponding to the node breakup, $w_N$, can be estimated as 
\begin{equation} \label{j:eq4}
w_{N}=\frac{V_{N}}{V_{0}}=\frac{V_{N}}{V_{D}}\frac{V_{D}}{V_{0}},
\end{equation}
where $V_D$ is the disk volume. For bag breakup, $V_{D}/V_{0}$ is approximately equal to unity because the entire initial volume of the drop is converted into the disk (i.e., there is no undeformed core). The ratio $V_{N}/V_{D}$ indicates the volume fraction of the nodes relative to the disk, which is about $0.4$ as considered by \cite{jackiw2022prediction}.

The volume weight of the rim, $w_R$, can be evaluated from the expression given by \citep{jackiw2021aerodynamic}
\begin{equation} \label{j:eq5}
w_{R}=\frac{V_{R}}{V_{0}}=\frac{3\pi }{2}\left [ \left ( \frac{2R_{i}}{d_{0}} \right )\left ( \frac{h_{i}}{d_{0}} \right )^{2}-\left ( \frac{h_{i}}{d_{0}} \right )^{3} \right ],
\end{equation}
where $h_i$ denotes the disk thickness and $R_i$ is the major radius of the rim. The value of $h_i$ and $R_i$ can be evaluated as \citep{jackiw2022prediction}
\begin{equation} \label{j:eq6}
\frac{h_{i}}{d_{0}}=\frac{4}{\We_{r}+5\left ( \frac{2R_{i}}{d_{0}} \right )-4\left ( \frac{d_{0}}{2R_{i}} \right )}-\frac{1}{20},
\end{equation}
and 
\begin{equation} \label{j:eq7}
\frac{2R_{i}}{d_{0}}=1.63-2.88e^{({-0.312\We})}.
\end{equation}
Here, $\We_{r}=\rho \dot{R}^{2}{d_{0}}/\sigma $ is the rim Weber number (which represents the competition between the radial momentum induced at droplet periphery and restoring surface tension of the stable droplet). The term $\dot{R}=U \sqrt{\rho_{a}/\rho}$ represents the constant radial expansion rate of a droplet \citep{marcotte2019density}. Here, $U$ is the average velocity of the airstream. 

The volume weight of the bag, $w_B$ is given by
\begin{equation} \label{j:eq8}
w_B=\frac{V_{B}}{V_{0}}=1-\frac{V_{N}}{V_{0}}-\frac{V_{R}}{V_{0}}.
\end{equation}
The method described above only provides volume weights for each mode. To obtain individual distributions as well as an overall distribution, characteristic droplet sizes must be determined for each mode. We, therefore, address the estimation of characteristic sizes for each mode in the following sections.

\subsubsection{Node droplet sizes ($d_N$)}
The nodes are formed on the rim due to the Rayleigh-Taylor (RT) instability \citep{zhao2010morphological} as the lighter fluid (air phase) pushes the heavier fluid (liquid phase). The droplet size ($d_N$) resulting from the breakup of nodes based on the RT instability theory is given by \citep{jackiw2022prediction} 
\begin{equation} \label{j:eq9}
\frac{d_{N}}{d_{0}}=\left [ \frac{3}{2}\left ( \frac{h_{i}}{d_{0}} \right )^{2}\frac{\lambda_{RT} }{d_{0}}n \right ]^{1/3},
\end{equation}
where $n=V_{N}/V_{D}$ represents the volume fraction of the nodes relative to the disk. \cite{jackiw2022prediction} estimated that the minimum, mean and maximum values of $n$ are 0.2, 0.4, and 1, respectively. The three characteristics sizes of node droplets can be determined using these three values of $n$.  The number-based mean and standard deviation for the breakup of the nodes are evaluated from the above-mentioned three characteristic sizes. In Eq. (\ref{j:eq9}), the maximum susceptible wavelength of the RT instability, $\lambda_{RT}=2\pi\sqrt{3\sigma/\rho a}$, wherein $a=\frac{3}{4}C_{D}\frac{U^{2}}{d_{0}}\frac{\rho_{a} }{\rho}\left ({D_{max}/d_{0}} \right )^{2}$ is the acceleration of the deforming droplet. As suggested by \cite{zhao2010morphological}, the drag coefficient ($C_{D}$) of the disk shape droplet is about 1.2 and the extent of droplet deformation, ${D_{max} / d_0}={2 / (1+\exp{(-0.0019 \We^{2.7})})}$.
 
\subsubsection{Rim droplet sizes ($d_R$)}
The satellite droplets associated with the rim breakup are due to the Rayleigh-Plateau instability, the receding rim instability \citep{jackiw2022prediction}, and the nonlinear instability of liquid ligaments near the pinch-off point.

The satellite droplets size ($d_R$) due to the Rayleigh-Plateau instability mechanism is 
\begin{equation} \label{j:eq14}
\frac{d_{R}}{d_{0}}=1.89\frac{h_{f}}{d_{0}}.
\end{equation} 
Here, $h_{f}$ is the final rim thickness, which is given by
\begin{equation} \label{j:eq15}
\frac{h_{f}}{d_{0}}=\frac{h_{i}}{d_{0}}\sqrt{\frac{R_{i}}{R_f}},
\end{equation}
where $R_f$ is the bag radius at the time of its burst. In case of swirl flow, $R_f$ can be evaluated as \citep{kirar2022experimental}
\begin{equation} \label{j:eq16}
R_f=\frac{d_{0}}{2\eta} \left [ 2e^{\tau ^{\prime}\sqrt{p}}+\left ( \frac{\sqrt{p}}{\sqrt{q}}-1 \right )e^{-\tau ^{\prime}\sqrt{q}}-\left ( \frac{\sqrt{p}}{\sqrt{q}}+1 \right )e^{\tau ^{\prime}\sqrt{q}} \right ],
\end{equation}   
where $\eta = f^{2}-120/\We$, $p=f^{2}-96/\We$ and $q=24/\We$.  \cite{kirar2022experimental} observed that the values of the stretching factor for the no-swirl case, $Sw=0.47$ and $Sw=0.82$ are $f=2.82$, $3.04$ and $3.41$, respectively. In Eq. (\ref{j:eq16}), the dimensionless time, $\tau^{\prime}=t_{b}/t_{d}$, wherein $t_b$ and $t_d$ are the bursting time and characteristic deformation time, respectively. They are given by \citep{jackiw2022prediction}
\begin{equation} \label{j:eq17}
t_{b}=\frac{\left [ \left ( \frac{2R_{i}}{d_{0}} \right )-2\left ( \frac{h_{i}}{d_{0}} \right ) \right ]}{\frac{2\dot{R}}{d_{0}}}\left [ -1+\sqrt{1+C\frac{8t_{d}}{\sqrt{3\We}}\frac{\frac{2\dot{R}}{d_{0}}}{\left [ \left ( \frac{2R_{i}}{d_{0}} \right )-2\left ( \frac{h_{i}}{d_{0}} \right ) \right ]}\sqrt{\frac{V_{B}}{V_{0}}}} \right ],
\end{equation} 
where $C=9.4$ \citep{jackiw2022prediction} and 
\begin{equation} \label{j:eq18}
t_{d}=\frac{d_0}{U}\sqrt{\frac{\rho }{\rho _{a}}}.
\end{equation}
The second mechanism responsible for the rim breakup is the receding rim instability \citep{jackiw2022prediction}, which results in a droplet size ($d_{rr}$) as given by
\begin{equation} \label{j:eq19}
\frac{d_{rr}}{d_{0}}=\left [ \frac{3}{2}\left ( \frac{h_{f}}{d_{0}} \right )^{2}\frac{\lambda _{rr}}{d_{0}} \right ]^{1/3},
\end{equation}
where $\lambda _{rr}$ is the wavelength of the receding rim instability, which is given by $\lambda _{rr}=4.5b_{rr}$. Here, $b_{rr}=\sqrt{\sigma /\rho a_{rr}}$ is the receding rim thickness, $a_{rr}=U_{rr}^{2}/R_{f}$ being the acceleration of the receding rim \citep{wang2018universal}, and  $U_{rr}$ is the receding rim velocity, which is measured experimentally.

The rim breakup occurs due to the nonlinear instability of liquid ligaments near the  pinch-off point. To obtained the characteristic size associated with this mechanism, we need to consider both the Rayleigh-Plateau and receding rim instabilities, which are given by \citep{keshavarz2020rotary}
\begin{equation} \label{j:eq20}
d_{sat,R}=\frac{d_R}{\sqrt{2+3Oh_{R}/\sqrt{2}}} ~~{\rm and}
\end{equation}   
\begin{equation} \label{j:eq21}
d_{sat,{rr}}=\frac{d_{rr}}{\sqrt{2+3Oh_{R}/\sqrt{2}}},
\end{equation}
respectively. Here, $Oh_{R}=\mu /\sqrt{\rho h_{f}^{3}\sigma }$ is the Ohnesorge number based on the final rim thickness. The characteristic sizes given in Eqs. (\ref{j:eq14}), (\ref{j:eq19}), (\ref{j:eq20}), and (\ref{j:eq21}) are used to evaluate number-based mean and standard deviation for the rim breakup.

\subsubsection{Bag droplet sizes ($d_{B}$)}
The factors that are responsible for the droplet size distribution due to rapture of bag-film are the minimum bag thickness, the receding rim thickness ($b_{rr}$), the Rayleigh-Plateu instability, and nonlinear instability of liquid ligaments. These factors lead to four characteristic sizes of the satellite droplets, which are given by \citep{jackiw2022prediction}
\begin{equation} \label{j:eq21_new}
 d_{B}=h_{min}, 
   \end{equation}  
   \begin{equation} \label{j:eq22}
 d_{rr,B}=b_{rr},
   \end{equation}  
    \begin{equation} \label{j:eq23}
 d_{RP,B}=1.89 b_{rr},
   \end{equation}  
   \begin{equation} \label{j:eq24}
d_{sat,B}=\frac{d_{RP,B}}{\sqrt{2+3Oh_{rr}/\sqrt{2}}}.
   \end{equation} 
\cite{jackiw2022prediction} found that $h_{min} = \pm 2.3$ $\mu \textrm{m}$ for the no-swirl case. Here, $Oh_{rr}=\mu /\sqrt{\rho b_{rr}^{3}\sigma }$ is the Ohnesorge number based on receding rim thickness, $b_{rr}$. These characteristic sizes are used to estimate the number mean and standard deviation associated with the bag fragmentation mode. 

\begin{figure}
\centering
\includegraphics[width=0.7\textwidth]{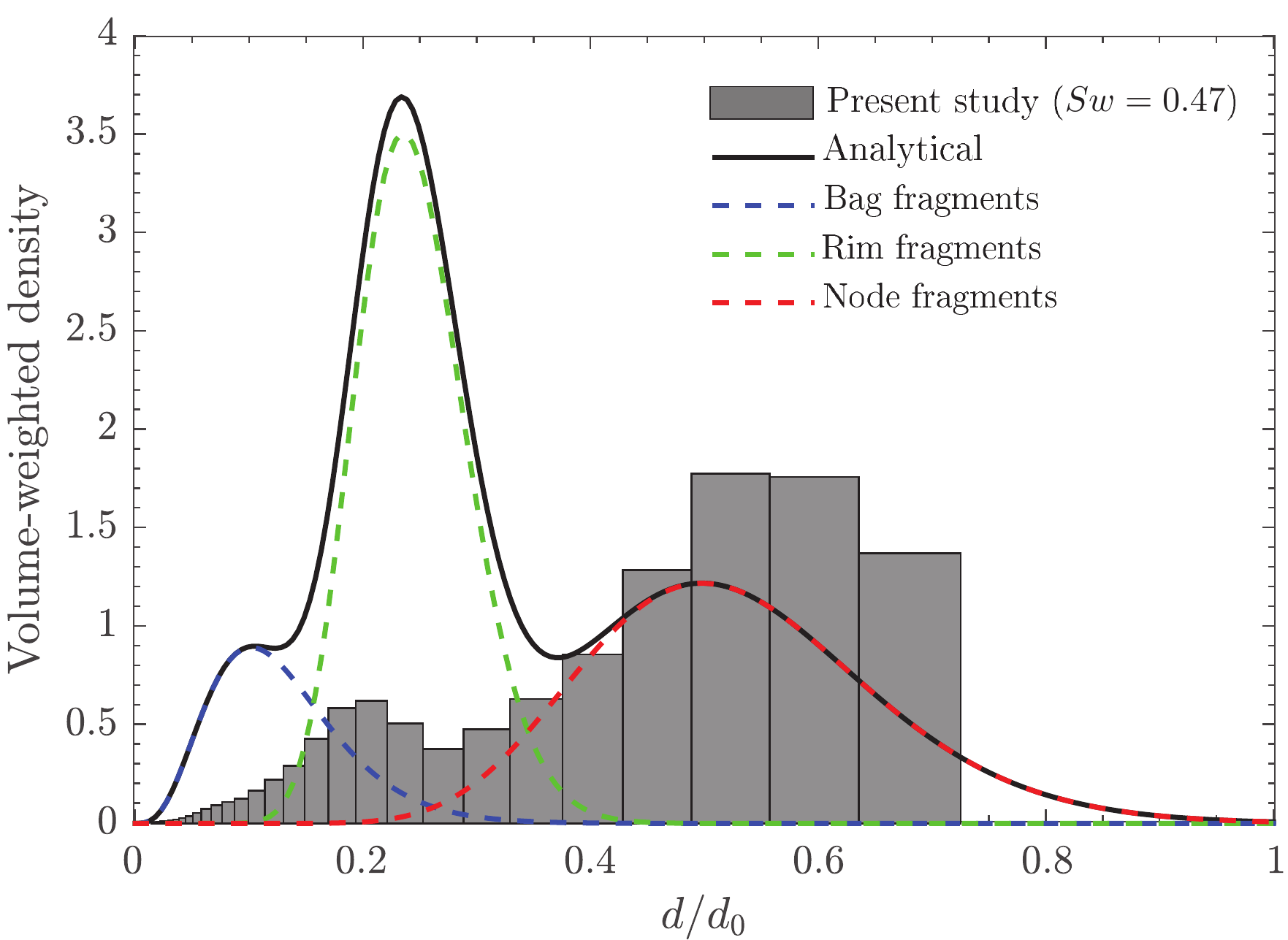}
\caption{Volume-weighted density of all fragments at $\tau=0.75$ for $Sw=0.47$ and $\We=12.1$. This depicts a bi-modal distribution.}
\label{fig9}
\end{figure}

\begin{figure}
\centering
\includegraphics[width=0.7\textwidth]{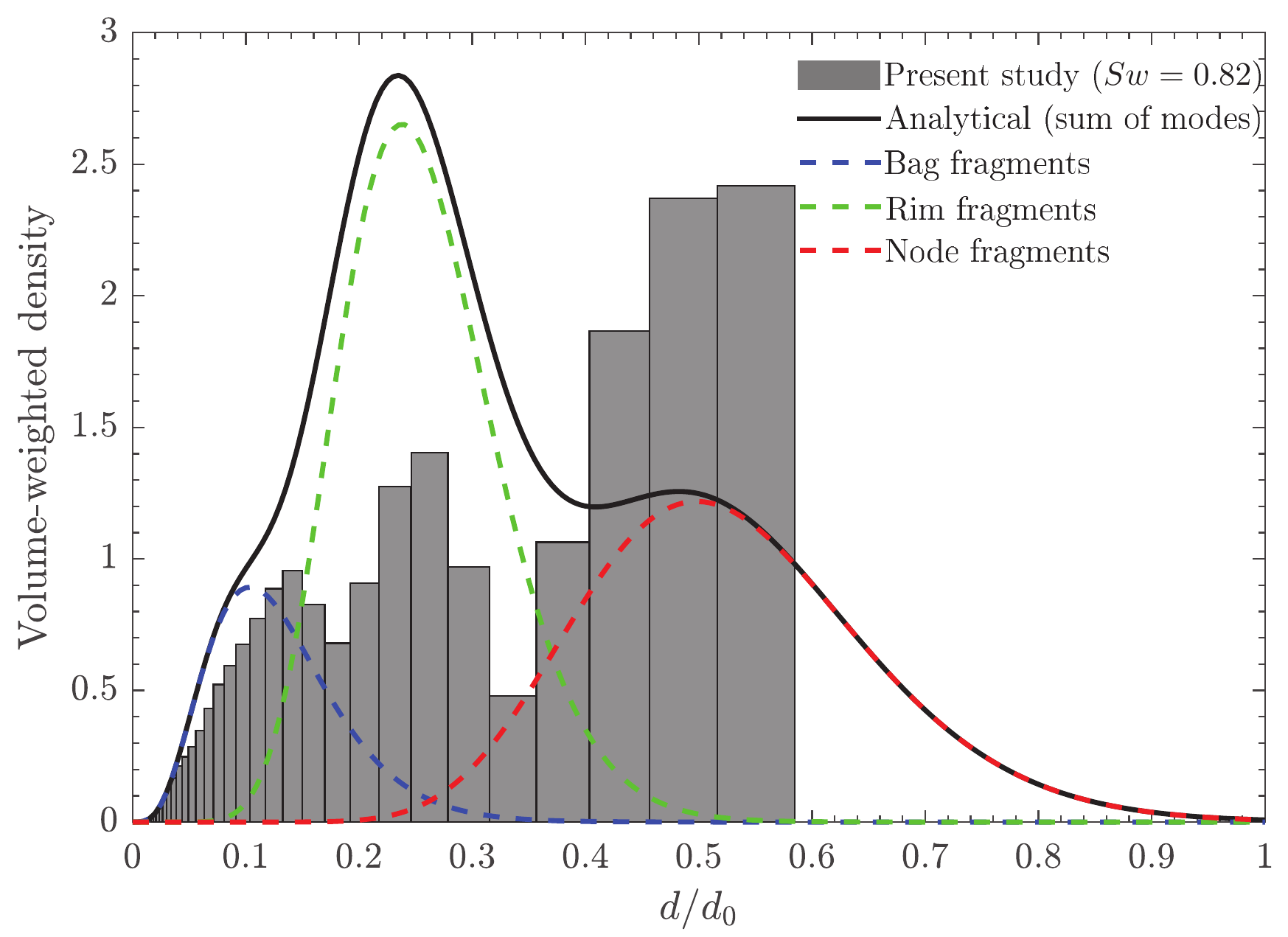}
\caption{Comparison of the volume-weighted density obtained from our experiments with the theoretical prediction (\cite{jackiw2022prediction}) at $\tau=0.66$ for $Sw=0.82$ and $\We=12.1$. Note the multi-modal distribution is likely due to fragments from the bag, the rim, and the node.}
\label{fig10}
\end{figure}

Figure \ref{fig9} shows the volume probability density of droplets for the low swirl ($Sw=0.47$) case at a typical instant, $\tau=0.75$ (this corresponds to the instant at which the fragmentation ceases). The uncertainty in the droplet size distributions from three experimental repetitions is presented in \ks{figure S5} of the supplementary material. The corresponding holography and shadowgraphy images are depicted in figure \ref{fig5}. As evident from the size distribution shown in figure \ref{fig9}, there are two distinct modes for the low swirl case ($Sw=0.47$). It is to be noted that, for retracting bag breakup, the bag fragmentation does not contribute to the size distribution. Therefore, only two geometries, the rim and node, contribute to the overall size distribution. In this case, due to the capillary instability, the breakup of the rim generates smaller droplets with a distribution peak at $d/d_0 \approx 0.20$, while the node breakup due to the Rayleigh-Taylor instability contributes to the droplet size distribution with a peak at $d/d_0 \approx 0.55$. Despite the fact that the rim breakup produces more number of small droplets (as seen in figure \ref{fig5} at $\tau=0.75$) than the node breakup, their contribution to the volume-weighted distribution is smaller since the volume is proportional to the cube of the droplet diameter. To the best of the author's knowledge, this is the first experimental study that reveals a  bi-modal distribution for the retracting bag breakup phenomenon. In order to theoretical predict the size distribution, we use the aforementioned model \citep{jackiw2022prediction} by considering bag, rim and node breakup mechanisms. However, in this scenario, the bag breakup does not produce the satellite droplets, and there are additional mechanisms present, such as total bag retraction and partial trapping of the rim in the wake zone of the airstream. We found that while the model reasonably predicts node breakup, it could not predict the size distribution for the bag and rim breakup processes.

In figure \ref{fig10}, we compare the analytical prediction of the combined distribution with the experimental results of the present study for the high swirl strength. \ks{Figure S6} in the supplementary material presents the uncertainty in the droplet size distributions obtained using three experimental repetitions. We observe that, unlike the low swirl ($Sw=0.47$), the high swirl ($Sw=0.82$) exhibits three distinct modes in the size distribution obtained experimentally as shown in figure \ref{fig10}. As explained earlier, for the high swirl case, the entire disk remains in the high shear zone. Therefore, the disk undergoes the normal bag breakup process, which involves the breakup of inflated bag, rim, and nodes. As a result, all these processes contribute to the overall size distribution. The first peak at $d/d_0 \approx 0.14$ in figure \ref{fig10} is attributed to the bag breakup phenomenon. The second peak observed at $d/d_0 \approx 0.24$ is due to the fragmentation of the rim, and the third peak at $d/d_0 \approx 0.45$ is due to the nodes breakup. As explained above, we have analytically calculated the characteristic sizes of each mode (bag, rim, and nodes) and then evaluated the individual size distribution of each mode separately. The theoretically predicted combined multi-modal distribution and that for the individual modes are shown in figure \ref{fig10}. The solid line indicates the sum of all the modes (bag+rim+nodes) of the size distribution, while the dashed lines represent the individual modes. These results reveal that the contributions of the bag and rim to the volume-weight size distributions are over-predicted, whereas the contribution to node breakup is under-estimated. However, the theoretically predicted characteristic sizes are in reasonable agreement with the experimental results. Therefore, this study demonstrates that the mode shapes and characteristic sizes can be estimated even for the bag breakup of a droplet under a swirl flow. Based on the current experimental data for the high swirl case, if we consider all the droplets larger than droplet size $d/d_{0}=0.16$ to originate from rim and nodes, the total volume fraction of rim and nodes droplets is $78\%$ (approximately). For the droplet fragmentation in an airstream without a swirl, \cite{guildenbecher2017characterization,gao2013uncertainty} have reported the total volume fraction of the droplets resulting from rim and nodes breakups to be $\sim 88\%$ and $\sim 90\%$, respectively.

\section{Concluding remarks}
\label{sec:conc}
The shape and size distribution of raindrops is one of the important factors in rainfall modelling, which is influenced by several microphysical processes, such as fragmentation, coalescence and phase change. In the present study, we examine the size distribution of satellite droplets produced by the interaction of a freely falling water droplet with a swirl airstream of different strengths using shadowgraphy and digital in-line holography techniques. The Weber number, which measures the aerodynamic force exerted on a droplet by the airstream, is considered to be low enough that the droplet undergoes vibrational breakup, producing only a few satellite droplets of comparable sizes in the no-swirl airstream. As we increase the swirl strength, the water droplet experiences the retracting bag breakup and normal breakup phenomena for $Sw=0.47$ (low swirl strength) and $Sw=0.82$ (high swirl strength), respectively. While \cite{kirar2022experimental} observed similar morphological changes for an ethanol droplet, they did not examine the size distribution of the satellite droplets following fragmentation, which is the subject of the current investigation. The digital in-line holography employed to obtain the size distributions of satellite droplets is a deep-learning-based image processing method that has recently emerged as a powerful tool for capturing three-dimensional information with high spatial resolution. Therefore, it is a more appropriate approach to be employed in the current study for the analysis of the size distribution of satellite droplets produced by fragmentation of a water droplet in a swirling airstream.

We observe that the number mean diameter of the satellite droplets increases over time due to the decrease in the relative velocity between the air and liquid bulk during the fragmentation process. In the high swirl case ($Sw=0.82$), the disintegration of the bag, rim, and nodes leads to smaller satellite droplets, whereas in the low swirl scenario ($Sw=0.47$), the fragmentation of the rim and nodes leads to larger satellite droplets. The bag rupture does not contribute to the size distribution for the low swirl case. The temporal variations of the Sauter mean diameter reveals that, for a given aerodynamic force, a high swirl strength creates more surface area (and thus more surface energy) than a low swirl strength.

A statistical theoretical analysis, similar to that of \cite{Villermaux2009single} originally developed for straight airstream without swirl, is also carried out to determine the size distribution of droplets for various swirl strengths. Despite not accounting for all breakup modes, the statistical model predicts the experimental result for small droplets but differs from the results for large droplets when the swirl strength is low. In the case of low swirl number, additional mechanisms such as entrapment of disk in the wake zone of the swirler and absence of bag fragmentation results in larger droplets. 

Furthermore, we have analytically evaluated the combined multi-modal distribution that accounts for nodes, rim, and bag breakup modes of a droplet in a swirl airflow. In sharp contrast to the earlier studies on the fragmentation of a droplet in a straight airstream without a swirl that observed a monomodal size distribution, we observed bimodal and multi-model distributions for low and high swirl strengths, respectively. We observe two distinct modes for the low swirl scenario $(Sw = 0.47)$ due to the rim breakup driven by the capillary instability, which results in smaller droplets with a symmetric distribution, and the subsequent node breakup, which generates an asymmetrical size distribution. For $Sw = 0.47$, we found that the deviations between experiments and the theoretical model predictions in the characteristic size of the droplets owing to the rim and node breakups are about 30\% and 16\%, respectively. In contrast, in the high swirl scenario $(Sw = 0.82$), we observe three peaks in the size distribution due to the breakup of the bag, rim and nodes. Although the theoretically evaluated characteristic sizes are in reasonable agreement with the experimental results, the volume-weighted contributions of the bag and rim are over-predicted, whereas the contribution to node breakup is under-predicted. Quantitatively, for $Sw = 0.82$, the deviations between experiments and the theoretical model predictions in the characteristic size of the droplets due to the bag, rim and node fragmentation are about 25\%, 4\% and 4\%, respectively. Nevertheless, the present study is a first attempt to estimate the size distribution of satellite droplets under a swirl airstream, which is observed in many natural and industrial applications.\\


\noindent{\bf Declaration of Interests:} The authors report no conflict of interest. \\
\\
\noindent{\bf Acknowledgement:} {L.D.C. and K.C.S. thank the Science \& Engineering Research Board, India for their financial support through grants SRG/2021/001048 and CRG/2020/000507, respectively. We thank IIT Hyderabad for the financial support through grant IITH/CHE/F011/SOCH1. S.S.A. also thanks the PMRF Fellowship.}


\end{document}